\hsize=31pc 
\vsize=49pc 
\lineskip=0pt 
\parskip=0pt plus 1pt 
\hfuzz=1pt   
\vfuzz=2pt 
\pretolerance=2500 
\tolerance=5000 
\vbadness=5000 
\hbadness=5000 
\widowpenalty=500 
\clubpenalty=200 
\brokenpenalty=500 
\predisplaypenalty=200 
\voffset=-1pc 
\nopagenumbers      
\catcode`@=11 
\newif\ifams 
\amsfalse 
%
%
%
\newfam\bdifam 
\newfam\bsyfam 
\newfam\bssfam 
\newfam\msafam 
\newfam\msbfam 
\newif\ifxxpt    
\newif\ifxviipt  
\newif\ifxivpt   
\newif\ifxiipt   
\newif\ifxipt    
\newif\ifxpt     
\newif\ifixpt    
\newif\ifviiipt  
\newif\ifviipt   
\newif\ifvipt    
\newif\ifvpt     
%
%
\def\headsize#1#2{\def\headb@seline{#2}%
                \ifnum#1=20\def\HEAD{twenty}%
                           \def\smHEAD{twelve}%
                           \def\vsHEAD{nine}%
                           \ifxxpt\else\xdef\f@ntsize{\HEAD}%
                           \def\m@g{4}\def\s@ze{20.74}%
                           \loadheadfonts\xxpttrue\fi 
                           \ifxiipt\else\xdef\f@ntsize{\smHEAD}%
                           \def\m@g{1}\def\s@ze{12}%
                           \loadxiiptfonts\xiipttrue\fi 
                           \ifixpt\else\xdef\f@ntsize{\vsHEAD}%
                           \def\s@ze{9}%
                           \loadsmallfonts\ixpttrue\fi 
                      \else 
                \ifnum#1=17\def\HEAD{seventeen}%
                           \def\smHEAD{eleven}%
                           \def\vsHEAD{eight}%
                           \ifxviipt\else\xdef\f@ntsize{\HEAD}%
                           \def\m@g{3}\def\s@ze{17.28}%
                           \loadheadfonts\xviipttrue\fi 
                           \ifxipt\else\xdef\f@ntsize{\smHEAD}%
                           \loadxiptfonts\xipttrue\fi 
                           \ifviiipt\else\xdef\f@ntsize{\vsHEAD}%
                           \def\s@ze{8}%
                           \loadsmallfonts\viiipttrue\fi 
                      \else\def\HEAD{fourteen}%
                           \def\smHEAD{ten}%
                           \def\vsHEAD{seven}%
                           \ifxivpt\else\xdef\f@ntsize{\HEAD}%
                           \def\m@g{2}\def\s@ze{14.4}%
                           \loadheadfonts\xivpttrue\fi 
                           \ifxpt\else\xdef\f@ntsize{\smHEAD}%
                           \def\s@ze{10}%
                           \loadxptfonts\xpttrue\fi 
                           \ifviipt\else\xdef\f@ntsize{\vsHEAD}%
                           \def\s@ze{7}%
                           \loadviiptfonts\viipttrue\fi 
                \ifnum#1=14\else 
                \message{Header size should be 20, 17 or 14 point 
                              will now default to 14pt}\fi 
                \fi\fi\headfonts} 
%
%
\def\textsize#1#2{\def\textb@seline{#2}%
                 \ifnum#1=12\def\TEXT{twelve}%
                           \def\smTEXT{eight}%
                           \def\vsTEXT{six}%
                           \ifxiipt\else\xdef\f@ntsize{\TEXT}%
                           \def\m@g{1}\def\s@ze{12}%
                           \loadxiiptfonts\xiipttrue\fi 
                           \ifviiipt\else\xdef\f@ntsize{\smTEXT}%
                           \def\s@ze{8}%
                           \loadsmallfonts\viiipttrue\fi 
                           \ifvipt\else\xdef\f@ntsize{\vsTEXT}%
                           \def\s@ze{6}%
                           \loadviptfonts\vipttrue\fi 
                      \else 
                \ifnum#1=11\def\TEXT{eleven}%
                           \def\smTEXT{seven}%
                           \def\vsTEXT{five}%
                           \ifxipt\else\xdef\f@ntsize{\TEXT}%
                           \def\s@ze{11}%
                           \loadxiptfonts\xipttrue\fi 
                           \ifviipt\else\xdef\f@ntsize{\smTEXT}%
                           \loadviiptfonts\viipttrue\fi 
                           \ifvpt\else\xdef\f@ntsize{\vsTEXT}%
                           \def\s@ze{5}%
                           \loadvptfonts\vpttrue\fi 
                      \else\def\TEXT{ten}%
                           \def\smTEXT{seven}%
                           \def\vsTEXT{five}%
                           \ifxpt\else\xdef\f@ntsize{\TEXT}%
                           \loadxptfonts\xpttrue\fi 
                           \ifviipt\else\xdef\f@ntsize{\smTEXT}%
                           \def\s@ze{7}%
                           \loadviiptfonts\viipttrue\fi 
                           \ifvpt\else\xdef\f@ntsize{\vsTEXT}%
                           \def\s@ze{5}%
                           \loadvptfonts\vpttrue\fi 
                \ifnum#1=10\else 
                \message{Text size should be 12, 11 or 10 point 
                              will now default to 10pt}\fi 
                \fi\fi\textfonts} 
%
%
\def\smallsize#1#2{\def\smallb@seline{#2}%
                 \ifnum#1=10\def\SMALL{ten}%
                           \def\smSMALL{seven}%
                           \def\vsSMALL{five}%
                           \ifxpt\else\xdef\f@ntsize{\SMALL}%
                           \loadxptfonts\xpttrue\fi 
                           \ifviipt\else\xdef\f@ntsize{\smSMALL}%
                           \def\s@ze{7}%
                           \loadviiptfonts\viipttrue\fi 
                           \ifvpt\else\xdef\f@ntsize{\vsSMALL}%
                           \def\s@ze{5}%
                           \loadvptfonts\vpttrue\fi 
                       \else 
                 \ifnum#1=9\def\SMALL{nine}%
                           \def\smSMALL{six}%
                           \def\vsSMALL{five}%
                           \ifixpt\else\xdef\f@ntsize{\SMALL}%
                           \def\s@ze{9}%
                           \loadsmallfonts\ixpttrue\fi 
                           \ifvipt\else\xdef\f@ntsize{\smSMALL}%
                           \def\s@ze{6}%
                           \loadviptfonts\vipttrue\fi 
                           \ifvpt\else\xdef\f@ntsize{\vsSMALL}%
                           \def\s@ze{5}%
                           \loadvptfonts\vpttrue\fi 
                       \else 
                           \def\SMALL{eight}%
                           \def\smSMALL{six}%
                           \def\vsSMALL{five}%
                           \ifviiipt\else\xdef\f@ntsize{\SMALL}%
                           \def\s@ze{8}%
                           \loadsmallfonts\viiipttrue\fi 
                           \ifvipt\else\xdef\f@ntsize{\smSMALL}%
                           \def\s@ze{6}%
                           \loadviptfonts\vipttrue\fi 
                           \ifvpt\else\xdef\f@ntsize{\vsSMALL}%
                           \def\s@ze{5}%
                           \loadvptfonts\vpttrue\fi 
                 \ifnum#1=8\else\message{Small size should be 10, 9 or  
                            8 point will now default to 8pt}\fi 
                \fi\fi\smallfonts} 
\def\F@nt{\expandafter\font\csname} 
\def\Sk@w{\expandafter\skewchar\csname} 
\def\@nd{\endcsname} 
\def\@step#1{ scaled \magstep#1} 
\def\@half{ scaled \magstephalf} 
\def\@t#1{ at #1pt} 
%
%
\def\loadheadfonts{\bigf@nts 
\F@nt \f@ntsize bdi\@nd=cmmib10 \@t{\s@ze}%
\Sk@w \f@ntsize bdi\@nd='177 
\F@nt \f@ntsize bsy\@nd=cmbsy10 \@t{\s@ze}%
\Sk@w \f@ntsize bsy\@nd='60 
\F@nt \f@ntsize bss\@nd=cmssbx10 \@t{\s@ze}} 
%
%
\def\loadxiiptfonts{\bigf@nts 
\F@nt \f@ntsize bdi\@nd=cmmib10 \@step{\m@g}%
\Sk@w \f@ntsize bdi\@nd='177 
\F@nt \f@ntsize bsy\@nd=cmbsy10 \@step{\m@g}%
\Sk@w \f@ntsize bsy\@nd='60 
\F@nt \f@ntsize bss\@nd=cmssbx10 \@step{\m@g}} 
%
%
\def\loadxiptfonts{%
\font\elevenrm=cmr10 \@half 
\font\eleveni=cmmi10 \@half 
\skewchar\eleveni='177 
\font\elevensy=cmsy10 \@half 
\skewchar\elevensy='60 
\font\elevenex=cmex10 \@half 
\font\elevenit=cmti10 \@half 
\font\elevensl=cmsl10 \@half 
\font\elevenbf=cmbx10 \@half 
\font\eleventt=cmtt10 \@half 
\ifams\font\elevenmsa=msam10 \@half 
\font\elevenmsb=msbm10 \@half\else\fi 
\font\elevenbdi=cmmib10 \@half 
\skewchar\elevenbdi='177 
\font\elevenbsy=cmbsy10 \@half 
\skewchar\elevenbsy='60 
\font\elevenbss=cmssbx10 \@half} 
%
%
\def\loadxptfonts{%
\font\tenbdi=cmmib10 
\skewchar\tenbdi='177 
\font\tenbsy=cmbsy10  
\skewchar\tenbsy='60 
\ifams\font\tenmsa=msam10  
\font\tenmsb=msbm10\else\fi 
\font\tenbss=cmssbx10}%
%
%
\def\loadsmallfonts{\smallf@nts 
\ifams 
\F@nt \f@ntsize ex\@nd=cmex\s@ze 
\else 
\F@nt \f@ntsize ex\@nd=cmex10\fi 
\F@nt \f@ntsize it\@nd=cmti\s@ze 
\F@nt \f@ntsize sl\@nd=cmsl\s@ze 
\F@nt \f@ntsize tt\@nd=cmtt\s@ze} 
%
%
\def\loadviiptfonts{%
\font\sevenit=cmti7 
\font\sevensl=cmsl8 at 7pt 
\ifams\font\sevenmsa=msam7  
\font\sevenmsb=msbm7 
\font\sevenex=cmex7 
\font\sevenbsy=cmbsy7 
\font\sevenbdi=cmmib7\else 
\font\sevenex=cmex10 
\font\sevenbsy=cmbsy10 at 7pt 
\font\sevenbdi=cmmib10 at 7pt\fi 
\skewchar\sevenbsy='60 
\skewchar\sevenbdi='177 
\font\sevenbss=cmssbx10 at 7pt}%
%
%
\def\loadviptfonts{\smallf@nts 
\ifams\font\sixex=cmex7 at 6pt\else 
\font\sixex=cmex10\fi 
\font\sixit=cmti7 at 6pt} 
%
%
\def\loadvptfonts{%
\font\fiveit=cmti7 at 5pt 
\ifams\font\fiveex=cmex7 at 5pt 
\font\fivebdi=cmmib5 
\font\fivebsy=cmbsy5 
\font\fivemsa=msam5  
\font\fivemsb=msbm5\else 
\font\fiveex=cmex10 
\font\fivebdi=cmmib10 at 5pt 
\font\fivebsy=cmbsy10 at 5pt\fi 
\skewchar\fivebdi='177 
\skewchar\fivebsy='60 
\font\fivebss=cmssbx10 at 5pt} 
\def\bigf@nts{%
\F@nt \f@ntsize rm\@nd=cmr10 \@step{\m@g}%
\F@nt \f@ntsize i\@nd=cmmi10 \@step{\m@g}%
\Sk@w \f@ntsize i\@nd='177 
\F@nt \f@ntsize sy\@nd=cmsy10 \@step{\m@g}%
\Sk@w \f@ntsize sy\@nd='60 
\F@nt \f@ntsize ex\@nd=cmex10 \@step{\m@g}%
\F@nt \f@ntsize it\@nd=cmti10 \@step{\m@g}%
\F@nt \f@ntsize sl\@nd=cmsl10 \@step{\m@g}%
\F@nt \f@ntsize bf\@nd=cmbx10 \@step{\m@g}%
\F@nt \f@ntsize tt\@nd=cmtt10 \@step{\m@g}%
\ifams 
\F@nt \f@ntsize msa\@nd=msam10 \@step{\m@g}%
\F@nt \f@ntsize msb\@nd=msbm10 \@step{\m@g}\else\fi} 
\def\smallf@nts{%
\F@nt \f@ntsize rm\@nd=cmr\s@ze 
\F@nt \f@ntsize i\@nd=cmmi\s@ze  
\Sk@w \f@ntsize i\@nd='177 
\F@nt \f@ntsize sy\@nd=cmsy\s@ze 
\Sk@w \f@ntsize sy\@nd='60 
\F@nt \f@ntsize bf\@nd=cmbx\s@ze  
\ifams 
\F@nt \f@ntsize bdi\@nd=cmmib\s@ze  
\F@nt \f@ntsize bsy\@nd=cmbsy\s@ze  
\F@nt \f@ntsize msa\@nd=msam\s@ze  
\F@nt \f@ntsize msb\@nd=msbm\s@ze 
\else 
\F@nt \f@ntsize bdi\@nd=cmmib10 \@t{\s@ze}%
\F@nt \f@ntsize bsy\@nd=cmbsy10 \@t{\s@ze}\fi  
\Sk@w \f@ntsize bdi\@nd='177 
\Sk@w \f@ntsize bsy\@nd='60 
\F@nt \f@ntsize bss\@nd=cmssbx10 \@t{\s@ze}}%
%
%
\def\headfonts{%
\textfont0=\csname\HEAD rm\@nd         
\scriptfont0=\csname\smHEAD rm\@nd 
\scriptscriptfont0=\csname\vsHEAD rm\@nd 
\def\rm{\fam0\csname\HEAD rm\@nd 
\def\sc{\csname\smHEAD rm\@nd}}%
\textfont1=\csname\HEAD i\@nd          
\scriptfont1=\csname\smHEAD i\@nd 
\scriptscriptfont1=\csname\vsHEAD i\@nd 
\textfont2=\csname\HEAD sy\@nd         
\scriptfont2=\csname\smHEAD sy\@nd 
\scriptscriptfont2=\csname\vsHEAD sy\@nd 
\textfont3=\csname\HEAD ex\@nd         
\scriptfont3=\csname\smHEAD ex\@nd 
\scriptscriptfont3=\csname\smHEAD ex\@nd 
\textfont\itfam=\csname\HEAD it\@nd    
\scriptfont\itfam=\csname\smHEAD it\@nd 
\scriptscriptfont\itfam=\csname\vsHEAD it\@nd 
\def\it{\fam\itfam\csname\HEAD it\@nd 
\def\sc{\csname\smHEAD it\@nd}}%
\textfont\slfam=\csname\HEAD sl\@nd    
\def\sl{\fam\slfam\csname\HEAD sl\@nd 
\def\sc{\csname\smHEAD sl\@nd}}%
\textfont\bffam=\csname\HEAD bf\@nd    
\scriptfont\bffam=\csname\smHEAD bf\@nd 
\scriptscriptfont\bffam=\csname\vsHEAD bf\@nd 
\def\bf{\fam\bffam\csname\HEAD bf\@nd 
\def\sc{\csname\smHEAD bf\@nd}}%
\textfont\ttfam=\csname\HEAD tt\@nd    
\def\tt{\fam\ttfam\csname\HEAD tt\@nd}%
\textfont\bdifam=\csname\HEAD bdi\@nd  
\scriptfont\bdifam=\csname\smHEAD bdi\@nd 
\scriptscriptfont\bdifam=\csname\vsHEAD bdi\@nd 
\def\bdi{\fam\bdifam\csname\HEAD bdi\@nd}%
\textfont\bsyfam=\csname\HEAD bsy\@nd  
\scriptfont\bsyfam=\csname\smHEAD bsy\@nd 
\def\bsy{\fam\bsyfam\csname\HEAD bsy\@nd}%
\textfont\bssfam=\csname\HEAD bss\@nd  
\scriptfont\bssfam=\csname\smHEAD bss\@nd 
\scriptscriptfont\bssfam=\csname\vsHEAD bss\@nd 
\def\bss{\fam\bssfam\csname\HEAD bss\@nd}%
\ifams 
\textfont\msafam=\csname\HEAD msa\@nd  
\scriptfont\msafam=\csname\smHEAD msa\@nd 
\scriptscriptfont\msafam=\csname\vsHEAD msa\@nd 
\textfont\msbfam=\csname\HEAD msb\@nd  
\scriptfont\msbfam=\csname\smHEAD msb\@nd 
\scriptscriptfont\msbfam=\csname\vsHEAD msb\@nd 
\else\fi 
\normalbaselineskip=\headb@seline pt%
\setbox\strutbox=\hbox{\vrule height.7\normalbaselineskip  
depth.3\baselineskip width0pt}%
\def\sc{\csname\smHEAD rm\@nd}\normalbaselines\bf} 
%
%
\def\textfonts{%
\textfont0=\csname\TEXT rm\@nd         
\scriptfont0=\csname\smTEXT rm\@nd 
\scriptscriptfont0=\csname\vsTEXT rm\@nd 
\def\rm{\fam0\csname\TEXT rm\@nd 
\def\sc{\csname\smTEXT rm\@nd}}%
\textfont1=\csname\TEXT i\@nd          
\scriptfont1=\csname\smTEXT i\@nd 
\scriptscriptfont1=\csname\vsTEXT i\@nd 
\textfont2=\csname\TEXT sy\@nd         
\scriptfont2=\csname\smTEXT sy\@nd 
\scriptscriptfont2=\csname\vsTEXT sy\@nd 
\textfont3=\csname\TEXT ex\@nd         
\scriptfont3=\csname\smTEXT ex\@nd 
\scriptscriptfont3=\csname\smTEXT ex\@nd 
\textfont\itfam=\csname\TEXT it\@nd    
\scriptfont\itfam=\csname\smTEXT it\@nd 
\scriptscriptfont\itfam=\csname\vsTEXT it\@nd 
\def\it{\fam\itfam\csname\TEXT it\@nd 
\def\sc{\csname\smTEXT it\@nd}}%
\textfont\slfam=\csname\TEXT sl\@nd    
\def\sl{\fam\slfam\csname\TEXT sl\@nd 
\def\sc{\csname\smTEXT sl\@nd}}%
\textfont\bffam=\csname\TEXT bf\@nd    
\scriptfont\bffam=\csname\smTEXT bf\@nd 
\scriptscriptfont\bffam=\csname\vsTEXT bf\@nd 
\def\bf{\fam\bffam\csname\TEXT bf\@nd 
\def\sc{\csname\smTEXT bf\@nd}}%
\textfont\ttfam=\csname\TEXT tt\@nd    
\def\tt{\fam\ttfam\csname\TEXT tt\@nd}%
\textfont\bdifam=\csname\TEXT bdi\@nd  
\scriptfont\bdifam=\csname\smTEXT bdi\@nd 
\scriptscriptfont\bdifam=\csname\vsTEXT bdi\@nd 
\def\bdi{\fam\bdifam\csname\TEXT bdi\@nd}%
\textfont\bsyfam=\csname\TEXT bsy\@nd  
\scriptfont\bsyfam=\csname\smTEXT bsy\@nd 
\def\bsy{\fam\bsyfam\csname\TEXT bsy\@nd}%
\textfont\bssfam=\csname\TEXT bss\@nd  
\scriptfont\bssfam=\csname\smTEXT bss\@nd 
\scriptscriptfont\bssfam=\csname\vsTEXT bss\@nd 
\def\bss{\fam\bssfam\csname\TEXT bss\@nd}%
\ifams 
\textfont\msafam=\csname\TEXT msa\@nd  
\scriptfont\msafam=\csname\smTEXT msa\@nd 
\scriptscriptfont\msafam=\csname\vsTEXT msa\@nd 
\textfont\msbfam=\csname\TEXT msb\@nd  
\scriptfont\msbfam=\csname\smTEXT msb\@nd 
\scriptscriptfont\msbfam=\csname\vsTEXT msb\@nd 
\else\fi 
\normalbaselineskip=\textb@seline pt 
\setbox\strutbox=\hbox{\vrule height.7\normalbaselineskip  
depth.3\baselineskip width0pt}%
\everymath{}%
\def\sc{\csname\smTEXT rm\@nd}\normalbaselines\rm} 
%
%
\def\smallfonts{%
\textfont0=\csname\SMALL rm\@nd         
\scriptfont0=\csname\smSMALL rm\@nd 
\scriptscriptfont0=\csname\vsSMALL rm\@nd 
\def\rm{\fam0\csname\SMALL rm\@nd 
\def\sc{\csname\smSMALL rm\@nd}}%
\textfont1=\csname\SMALL i\@nd          
\scriptfont1=\csname\smSMALL i\@nd 
\scriptscriptfont1=\csname\vsSMALL i\@nd 
\textfont2=\csname\SMALL sy\@nd         
\scriptfont2=\csname\smSMALL sy\@nd 
\scriptscriptfont2=\csname\vsSMALL sy\@nd 
\textfont3=\csname\SMALL ex\@nd         
\scriptfont3=\csname\smSMALL ex\@nd 
\scriptscriptfont3=\csname\smSMALL ex\@nd 
\textfont\itfam=\csname\SMALL it\@nd    
\scriptfont\itfam=\csname\smSMALL it\@nd 
\scriptscriptfont\itfam=\csname\vsSMALL it\@nd 
\def\it{\fam\itfam\csname\SMALL it\@nd 
\def\sc{\csname\smSMALL it\@nd}}%
\textfont\slfam=\csname\SMALL sl\@nd    
\def\sl{\fam\slfam\csname\SMALL sl\@nd 
\def\sc{\csname\smSMALL sl\@nd}}%
\textfont\bffam=\csname\SMALL bf\@nd    
\scriptfont\bffam=\csname\smSMALL bf\@nd 
\scriptscriptfont\bffam=\csname\vsSMALL bf\@nd 
\def\bf{\fam\bffam\csname\SMALL bf\@nd 
\def\sc{\csname\smSMALL bf\@nd}}%
\textfont\ttfam=\csname\SMALL tt\@nd    
\def\tt{\fam\ttfam\csname\SMALL tt\@nd}%
\textfont\bdifam=\csname\SMALL bdi\@nd  
\scriptfont\bdifam=\csname\smSMALL bdi\@nd 
\scriptscriptfont\bdifam=\csname\vsSMALL bdi\@nd 
\def\bdi{\fam\bdifam\csname\SMALL bdi\@nd}%
\textfont\bsyfam=\csname\SMALL bsy\@nd  
\scriptfont\bsyfam=\csname\smSMALL bsy\@nd 
\def\bsy{\fam\bsyfam\csname\SMALL bsy\@nd}%
\textfont\bssfam=\csname\SMALL bss\@nd  
\scriptfont\bssfam=\csname\smSMALL bss\@nd 
\scriptscriptfont\bssfam=\csname\vsSMALL bss\@nd 
\def\bss{\fam\bssfam\csname\SMALL bss\@nd}%
\ifams 
\textfont\msafam=\csname\SMALL msa\@nd  
\scriptfont\msafam=\csname\smSMALL msa\@nd 
\scriptscriptfont\msafam=\csname\vsSMALL msa\@nd 
\textfont\msbfam=\csname\SMALL msb\@nd  
\scriptfont\msbfam=\csname\smSMALL msb\@nd 
\scriptscriptfont\msbfam=\csname\vsSMALL msb\@nd 
\else\fi 
\normalbaselineskip=\smallb@seline pt%
\setbox\strutbox=\hbox{\vrule height.7\normalbaselineskip  
depth.3\baselineskip width0pt}%
\everymath{}%
\def\sc{\csname\smSMALL rm\@nd}\normalbaselines\rm}%
\everydisplay{\indenteddisplay 
   \gdef\labeltype{\eqlabel}}%
%
%
\def\hexnumber@#1{\ifcase#1 0\or 1\or 2\or 3\or 4\or 5\or 6\or 7\or 8\or 
 9\or A\or B\or C\or D\or E\or F\fi} 
\edef\bffam@{\hexnumber@\bffam} 
\edef\bdifam@{\hexnumber@\bdifam} 
\edef\bsyfam@{\hexnumber@\bsyfam} 
\def\undefine#1{\let#1\undefined} 
\def\newsymbol#1#2#3#4#5{\let\next@\relax 
 \ifnum#2=\thr@@\let\next@\bdifam@\else 
 \ifams 
 \ifnum#2=\@ne\let\next@\msafam@\else 
 \ifnum#2=\tw@\let\next@\msbfam@\fi\fi 
 \fi\fi 
 \mathchardef#1="#3\next@#4#5} 
\def\mathhexbox@#1#2#3{\relax 
 \ifmmode\mathpalette{}{\m@th\mathchar"#1#2#3}%
 \else\leavevmode\hbox{$\m@th\mathchar"#1#2#3$}\fi} 

\def\bi#1{{\fam\bdifam\relax#1}} 
%
%
\ifams\input amsmacro\fi 
%
%
\newsymbol\bitGamma 3000 
\newsymbol\bitDelta 3001 
\newsymbol\bitTheta 3002 
\newsymbol\bitLambda 3003 
\newsymbol\bitXi 3004 
\newsymbol\bitPi 3005 
\newsymbol\bitSigma 3006 
\newsymbol\bitUpsilon 3007 
\newsymbol\bitPhi 3008 
\newsymbol\bitPsi 3009 
\newsymbol\bitOmega 300A 
\newsymbol\balpha 300B 
\newsymbol\bbeta 300C 
\newsymbol\bgamma 300D 
\newsymbol\bdelta 300E 
\newsymbol\bepsilon 300F 
\newsymbol\bzeta 3010 
\newsymbol\bfeta 3011 
\newsymbol\btheta 3012 
\newsymbol\biota 3013 
\newsymbol\bkappa 3014 
\newsymbol\blambda 3015 
\newsymbol\bmu 3016 
\newsymbol\bnu 3017 
\newsymbol\bxi 3018 
\newsymbol\bpi 3019 
\newsymbol\brho 301A 
\newsymbol\bsigma 301B 
\newsymbol\btau 301C 
\newsymbol\bupsilon 301D 
\newsymbol\bphi 301E 
\newsymbol\bchi 301F 
\newsymbol\bpsi 3020 
\newsymbol\bomega 3021 
\newsymbol\bvarepsilon 3022 
\newsymbol\bvartheta 3023 
\newsymbol\bvaromega 3024 
\newsymbol\bvarrho 3025 
\newsymbol\bvarzeta 3026 
\newsymbol\bvarphi 3027 
\newsymbol\bpartial 3040 
\newsymbol\bell 3060 
\newsymbol\bimath 307B 
\newsymbol\bjmath 307C 
\mathchardef\binfty "0\bsyfam@31 
\mathchardef\bnabla "0\bsyfam@72 
\mathchardef\bdot "2\bsyfam@01 
\mathchardef\bGamma "0\bffam@00 
\mathchardef\bDelta "0\bffam@01 
\mathchardef\bTheta "0\bffam@02 
\mathchardef\bLambda "0\bffam@03 
\mathchardef\bXi "0\bffam@04 
\mathchardef\bPi "0\bffam@05 
\mathchardef\bSigma "0\bffam@06 
\mathchardef\bUpsilon "0\bffam@07 
\mathchardef\bPhi "0\bffam@08 
\mathchardef\bPsi "0\bffam@09 
\mathchardef\bOmega "0\bffam@0A 
\mathchardef\itGamma "0100 
\mathchardef\itDelta "0101 
\mathchardef\itTheta "0102 
\mathchardef\itLambda "0103 
\mathchardef\itXi "0104 
\mathchardef\itPi "0105 
\mathchardef\itSigma "0106 
\mathchardef\itUpsilon "0107 
\mathchardef\itPhi "0108 
\mathchardef\itPsi "0109 
\mathchardef\itOmega "010A 
\mathchardef\Gamma "0000 
\mathchardef\Delta "0001 
\mathchardef\Theta "0002 
\mathchardef\Lambda "0003 
\mathchardef\Xi "0004 
\mathchardef\Pi "0005 
\mathchardef\Sigma "0006 
\mathchardef\Upsilon "0007 
\mathchardef\Phi "0008 
\mathchardef\Psi "0009 
\mathchardef\Omega "000A 
%
%
\newcount\firstpage  \firstpage=1  
\newcount\jnl                      
\newcount\secno                    
\newcount\subno                    
\newcount\subsubno                 
\newcount\appno                    
\newcount\tabno                    
\newcount\figno                    
\newcount\countno                  
\newcount\refno                    
\newcount\eqlett     \eqlett=97    
\newif\ifletter 
\newif\ifwide 
\newif\ifnotfull 
\newif\ifaligned 
\newif\ifnumbysec   
\newif\ifappendix 
\newif\ifnumapp 
\newif\ifssf 
\newif\ifppt 
\newdimen\t@bwidth 
\newdimen\c@pwidth 
\newdimen\digitwidth                    
\newdimen\argwidth                      
\newdimen\secindent    \secindent=5pc   
\newdimen\textind    \textind=16pt      
\newdimen\tempval                       
\newskip\beforesecskip 
\def\beforesecspace{\vskip\beforesecskip\relax} 
\newskip\beforesubskip 
\def\beforesubspace{\vskip\beforesubskip\relax} 
\newskip\beforesubsubskip 
\def\beforesubsubspace{\vskip\beforesubsubskip\relax} 
\newskip\secskip 
\def\secspace{\vskip\secskip\relax} 
\newskip\subskip 
\def\subspace{\vskip\subskip\relax} 
\newskip\insertskip 
\def\insertspace{\vskip\insertskip\relax} 
\def\sp@ce{\ifx\next*\let\next=\@ssf 
               \else\let\next=\@nossf\fi\next} 
\def\@ssf#1{\nobreak\secspace\global\ssftrue\nobreak} 
\def\@nossf{\nobreak\secspace\nobreak\noindent\ignorespaces} 
\def\subsp@ce{\ifx\next*\let\next=\@sssf 
               \else\let\next=\@nosssf\fi\next} 
\def\@sssf#1{\nobreak\subspace\global\ssftrue\nobreak} 
\def\@nosssf{\nobreak\subspace\nobreak\noindent\ignorespaces} 
\beforesecskip=24pt plus12pt minus8pt 
\beforesubskip=12pt plus6pt minus4pt 
\beforesubsubskip=12pt plus6pt minus4pt 
\secskip=12pt plus 2pt minus 2pt 
\subskip=6pt plus3pt minus2pt 
\insertskip=18pt plus6pt minus6pt%
\fontdimen16\tensy=2.7pt 
\fontdimen17\tensy=2.7pt 
%
%
\def\eqlabel{(\ifappendix\applett 
               \ifnumbysec\ifnum\secno>0 \the\secno\fi.\fi 
               \else\ifnumbysec\the\secno.\fi\fi\the\countno)} 
\def\seclabel{\ifappendix\ifnumapp\else\applett\fi 
    \ifnum\secno>0 \the\secno 
    \ifnumbysec\ifnum\subno>0.\the\subno\fi\fi\fi 
    \else\the\secno\fi\ifnum\subno>0.\the\subno 
         \ifnum\subsubno>0.\the\subsubno\fi\fi} 
\def\tablabel{\ifappendix\applett\fi\the\tabno} 
\def\figlabel{\ifappendix\applett\fi\the\figno} 
\def\gac{\global\advance\countno by 1} 
%
%
 
\def\vfootnote#1{\insert\footins\bgroup 
\interlinepenalty=\interfootnotelinepenalty 
\splittopskip=\ht\strutbox 
\splitmaxdepth=\dp\strutbox \floatingpenalty=20000 
\leftskip=0pt \rightskip=0pt \spaceskip=0pt \xspaceskip=0pt%
\noindent\smallfonts\rm #1\ \ignorespaces\footstrut\futurelet\next\fo@t} 
%
%
\def\endinsert{\egroup 
    \if@mid \dimen@=\ht0 \advance\dimen@ by\dp0 
       \advance\dimen@ by12\p@ \advance\dimen@ by\pagetotal 
       \ifdim\dimen@>\pagegoal \@midfalse\p@gefalse\fi\fi 
    \if@mid \insertspace \box0 \par \ifdim\lastskip<\insertskip 
    \removelastskip \penalty-200 \insertspace \fi 
    \else\insert\topins{\penalty100 
       \splittopskip=0pt \splitmaxdepth=\maxdimen  
       \floatingpenalty=0 
       \ifp@ge \dimen@=\dp0 
       \vbox to\vsize{\unvbox0 \kern-\dimen@}%
       \else\box0\nobreak\insertspace\fi}\fi\endgroup}    
%
%
%
\def\ind{\hbox to \secindent{\hfill}} 
%
%

%
%
\def\lo#1{\llap{${}#1{}$}} 
%
%
\def\indeqn#1{\alignedfalse\displ@y\halign{\hbox to \displaywidth 
    {$\ind\@lign\displaystyle##\hfil$}\crcr #1\crcr}} 
%
%
\def\indalign#1{\alignedtrue\displ@y \tabskip=0pt  
  \halign to\displaywidth{\ind$\@lign\displaystyle{##}$\tabskip=0pt 
    &$\@lign\displaystyle{{}##}$\hfill\tabskip=\centering 
    &\llap{$\@lign\hbox{\rm##}$}\tabskip=0pt\crcr 
    #1\crcr}} 
\def\fl{{\hskip-\secindent}} 
\def\indenteddisplay#1$${\indispl@y{#1 }} 
\def\indispl@y#1{\disptest#1\eqalignno\eqalignno\disptest} 
\def\disptest#1\eqalignno#2\eqalignno#3\disptest{%
    \ifx#3\eqalignno 
    \indalign#2%
    \else\indeqn{#1}\fi$$} 
%
%
 
%
%
 
%
%
 
%
%
 
%
%

\def\ns{\noalign{\vskip-3pt}}

%
 
%
%
\def\bhbar{\rlap{\kern1pt\raise.4ex\hbox{\bf\char'40}}\bi{h}} 

\def\dash{---{}--- }

\def\etal{{\it et al\/}\ } 
\def\frac#1#2{{#1\over#2}} 
\ifams 
\def\lap{\lesssim} 
\def\gap{\gtrsim}

\let\geq=\geqslant 
\else

\def\gap{\;\lower3pt\hbox{$\buildrel > \over \sim$}\;}%
\def\lap{\;\lower3pt\hbox{$\buildrel < \over \sim$}\;}\fi 
 
\chardef\ii="10 
\def\tqs{\hbox to 25pt{\hfil}}

\def\Bbbone{1\kern-.22em {\rm l}} 
%
%
\def\rp{\raise8pt\hbox{$\scriptstyle\prime$}} 
%
%
%
%

%
%
\def\[#1\]{\setbox0=\hbox{$\dsty#1$}\argwidth=\wd0 
    \setbox0=\hbox{$\left[\box0\right]$}\advance\argwidth by -\wd0 
    \left[\kern.3\argwidth\box0\kern.3\argwidth\right]} 
%
%
\def\lsb#1\rsb{\setbox0=\hbox{$#1$}\argwidth=\wd0 
    \setbox0=\hbox{$\left[\box0\right]$}\advance\argwidth by -\wd0 
    \left[\kern.3\argwidth\box0\kern.3\argwidth\right]} 
%
 
%
%
 
%
\def\pt(#1){({\it #1\/})} 
\let\dsty=\displaystyle

%
%
\def\reactions#1{\vskip 12pt plus2pt minus2pt%
\vbox{\hbox{\kern\secindent\vrule\kern12pt%
\vbox{\kern0.5pt\vbox{\hsize=24pc\parindent=0pt\smallfonts\rm NUCLEAR  
REACTIONS\strut\quad #1\strut}\kern0.5pt}\kern12pt\vrule}}} 
%
%
\def\slashchar#1{\setbox0=\hbox{$#1$}\dimen0=\wd0%
\setbox1=\hbox{/}\dimen1=\wd1%
\ifdim\dimen0>\dimen1%
\rlap{\hbox to \dimen0{\hfil/\hfil}}#1\else                                         
\rlap{\hbox to \dimen1{\hfil$#1$\hfil}}/\fi} 
%
%
\def\textindent#1{\noindent\hbox to \parindent{#1\hss}\ignorespaces} 
%
%
\def\opencirc{\raise1pt\hbox{$\scriptstyle{\bigcirc}$}} 
 
\ifams 
\def\opensqr{\hbox{$\square$}} 
 
\def\opentridown{\hbox{$\triangledown$}}

\else 
\def\opensqr{\vbox{\hrule height.4pt\hbox{\vrule width.4pt height3.5pt 
    \kern3.5pt\vrule width.4pt}\hrule height.4pt}} 
 
\def\opentridown{\raise1pt\hbox{$\scriptstyle\bigtriangledown$}}

\fi

%
%
\def\m@th{\mathsurround=0pt} 
%
%
\def\cases#1{%
\left\{\,\vcenter{\normalbaselines\openup1\jot\m@th%
     \ialign{$\displaystyle##\hfil$&\rm\tqs##\hfil\crcr#1\crcr}}\right.}%
%
%
\def\oldcases#1{\left\{\,\vcenter{\normalbaselines\m@th 
    \ialign{$##\hfil$&\rm\quad##\hfil\crcr#1\crcr}}\right.} 
%
%
\def\numcases#1{\left\{\,\vcenter{\baselineskip=15pt\m@th%
     \ialign{$\displaystyle##\hfil$&\rm\tqs##\hfil 
     \crcr#1\crcr}}\right.\hfill 
     \vcenter{\baselineskip=15pt\m@th%
     \ialign{\rlap{$\phantom{\displaystyle##\hfil}$}\tabskip=0pt&\en 
     \rlap{\phantom{##\hfil}}\crcr#1\crcr}}} 
\def\ptnumcases#1{\left\{\,\vcenter{\baselineskip=15pt\m@th%
     \ialign{$\displaystyle##\hfil$&\rm\tqs##\hfil 
     \crcr#1\crcr}}\right.\hfill 
     \vcenter{\baselineskip=15pt\m@th%
     \ialign{\rlap{$\phantom{\displaystyle##\hfil}$}\tabskip=0pt&\enpt 
     \rlap{\phantom{##\hfil}}\crcr#1\crcr}}\global\eqlett=97 
     \global\advance\countno by 1} 
%
%
\def\eq(#1){\ifaligned\@mp(#1)\else\hfill\llap{{\rm (#1)}}\fi} 
\def\ceq(#1){\ns\ns\ifaligned\@mp\fi\eq(#1)\cr\ns\ns} 
\def\eqpt(#1#2){\ifaligned\@mp(#1{\it #2\/}) 
                    \else\hfill\llap{{\rm (#1{\it #2\/})}}\fi} 
\let\eqno=\eq 
%
%
\countno=1 
 
\def\aleq{&\rm(\ifappendix\applett 
               \ifnumbysec\ifnum\secno>0 \the\secno\fi.\fi 
               \else\ifnumbysec\the\secno.\fi\fi\the\countno} 
\def\noaleq{\hfill\llap\bgroup\rm(\ifappendix\applett 
               \ifnumbysec\ifnum\secno>0 \the\secno\fi.\fi 
               \else\ifnumbysec\the\secno.\fi\fi\the\countno} 
\def\@mp{&} 
\def\en{\ifaligned\aleq)\else\noaleq)\egroup\fi\gac} 
\def\cen{\ns\ns\ifaligned\@mp\fi\en\cr\ns\ns} 
\def\enpt{\ifaligned\aleq{\it\char\the\eqlett})\else 
    \noaleq{\it\char\the\eqlett})\egroup\fi 
    \global\advance\eqlett by 1} 
\def\endpt{\ifaligned\aleq{\it\char\the\eqlett})\else 
    \noaleq{\it\char\the\eqlett})\egroup\fi 
    \global\eqlett=97\gac} 
%
%

\def\JPA{{\it J. Phys. A: Math. Gen.}} 
\def\JPC{{\it J. Phys. C: Solid State Phys.}}     

 

\def\RPP{{\it Rep. Prog. Phys.}}

%
%

\def\APNY{{\it Ann. Phys., NY\/}}

\def\JP{{\it J. Physique\/}}

\def\JPSJ{{\it J. Phys. Soc. Japan\/}}

\def\PL{{\it Phys. Lett.}} 
\def\PR{{\it Phys. Rev.}} 
\def\PRL{{\it Phys. Rev. Lett.}}

\def\RMP{{\it Rev. Mod. Phys.}}

\def\ZP{{\it Z. Phys.}} 
\headline={\ifodd\pageno{\ifnum\pageno=\firstpage\hfill 
   \else\rrhead\fi}\else\lrhead\fi} 
\def\rrhead{\textfonts\hskip\secindent\it 
    \shorttitle\hfill\rm\folio} 
\def\lrhead{\textfonts\hbox to\secindent{\rm\folio\hss}%
    \it\aunames\hss} 
\footline={\ifnum\pageno=\firstpage \hfill\textfonts\rm\folio\fi} 
\def\@rticle#1#2{\vglue.5pc 
    {\parindent=\secindent \bf #1\par} 
     \vskip2.5pc 
    {\exhyphenpenalty=10000\hyphenpenalty=10000 
     \baselineskip=18pt\raggedright\noindent 
     \headfonts\bf#2\par}\futurelet\next\sh@rttitle}%
\def\title#1{\gdef\shorttitle{#1} 
    \vglue4pc{\exhyphenpenalty=10000\hyphenpenalty=10000  
    \baselineskip=18pt  
    \raggedright\parindent=0pt 
    \headfonts\bf#1\par}\futurelet\next\sh@rttitle}  

\def\article#1#2{\gdef\shorttitle{#2}\@rticle{#1}{#2}}  
\def\review#1{\gdef\shorttitle{#1}%
    \@rticle{REVIEW \ifpbm\else ARTICLE\fi}{#1}} 
\def\topical#1{\gdef\shorttitle{#1}%
    \@rticle{TOPICAL REVIEW}{#1}} 
\def\comment#1{\gdef\shorttitle{#1}%
    \@rticle{COMMENT}{#1}} 
\def\note#1{\gdef\shorttitle{#1}%
    \@rticle{NOTE}{#1}} 
\def\prelim#1{\gdef\shorttitle{#1}%
    \@rticle{PRELIMINARY COMMUNICATION}{#1}} 
\def\letter#1{\gdef\shorttitle{Letter to the Editor}%
     \gdef\aunames{Letter to the Editor} 
     \global\lettertrue\ifnum\jnl=7\global\letterfalse\fi 
     \@rticle{LETTER TO THE EDITOR}{#1}} 
\def\sh@rttitle{\ifx\next[\let\next=\sh@rt 
                \else\let\next=\f@ll\fi\next} 
\def\sh@rt[#1]{\gdef\shorttitle{#1}} 
\def\f@ll{} 
\def\author#1{\ifletter\else\gdef\aunames{#1}\fi\vskip1.5pc 
    {\parindent=\secindent   
     \hang\textfonts   
     \ifppt\bf\else\rm\fi#1\par}   
     \ifppt\bigskip\else\smallskip\fi 
     \futurelet\next\@unames} 
\def\@unames{\ifx\next[\let\next=\short@uthor 
                 \else\let\next=\@uthor\fi\next} 
\def\short@uthor[#1]{\gdef\aunames{#1}} 
\def\@uthor{} 
\def\address#1{{\parindent=\secindent 
    \exhyphenpenalty=10000\hyphenpenalty=10000 
\ifppt\textfonts\else\smallfonts\fi\hang\raggedright\rm#1\par}%
    \ifppt\bigskip\fi} 
\def\jl#1{\global\jnl=#1} 
\jl{0}%
\def\journal{\ifnum\jnl=1 J. Phys.\ A: Math.\ Gen.\  
        \else\ifnum\jnl=2 J. Phys.\ B: At.\ Mol.\ Opt.\ Phys.\  
        \else\ifnum\jnl=3 J. Phys.:\ Condens. Matter\  
        \else\ifnum\jnl=4 J. Phys.\ G: Nucl.\ Part.\ Phys.\  
        \else\ifnum\jnl=5 Inverse Problems\  
        \else\ifnum\jnl=6 Class. Quantum Grav.\  
        \else\ifnum\jnl=7 Network\  
        \else\ifnum\jnl=8 Nonlinearity\ 
        \else\ifnum\jnl=9 Quantum Opt.\ 
        \else\ifnum\jnl=10 Waves in Random Media\ 
        \else\ifnum\jnl=11 Pure Appl. Opt.\  
        \else\ifnum\jnl=12 Phys. Med. Biol.\ 
        \else\ifnum\jnl=13 Modelling Simulation Mater.\ Sci.\ Eng.\  
        \else\ifnum\jnl=14 Plasma Phys. Control. Fusion\  
        \else\ifnum\jnl=15 Physiol. Meas.\  
        \else\ifnum\jnl=16 Sov.\ Lightwave Commun.\ 
        \else\ifnum\jnl=17 J. Phys.\ D: Appl.\ Phys.\ 
        \else\ifnum\jnl=18 Supercond.\ Sci.\ Technol.\ 
        \else\ifnum\jnl=19 Semicond.\ Sci.\ Technol.\ 
        \else\ifnum\jnl=20 Nanotechnology\ 
        \else\ifnum\jnl=21 Meas.\ Sci.\ Technol.\  
        \else\ifnum\jnl=22 Plasma Sources Sci.\ Technol.\  
        \else\ifnum\jnl=23 Smart Mater.\ Struct.\  
        \else\ifnum\jnl=24 J.\ Micromech.\ Microeng.\ 
   \else Institute of Physics Publishing\  
   \fi\fi\fi\fi\fi\fi\fi\fi\fi\fi\fi\fi\fi\fi\fi 
   \fi\fi\fi\fi\fi\fi\fi\fi\fi} 
\let\abs=\beginabstract 

\let\endabs=\endabstract 
\def\submitted{\ifppt\noindent\textfonts\rm Submitted to \journal\par 
     \bigskip\fi} 
\def\today{\number\day\ \ifcase\month\or 
     January\or February\or March\or April\or May\or June\or 
     July\or August\or September\or October\or November\or 
     December\fi\space \number\year} 
\def\date{\ifppt\noindent\textfonts\rm  
     Date: \today\par\goodbreak\bigskip\fi} 
%
%
\def\pacs#1{\ifppt\noindent\textfonts\rm  
     PACS number(s): #1\par\bigskip\fi} 
%
 
%
%
\def\section#1{\ifppt\ifnum\secno=0\eject\fi\fi 
    \subno=0\subsubno=0\global\advance\secno by 1 
    \gdef\labeltype{\seclabel}\ifnumbysec\countno=1\fi 
    \goodbreak\beforesecspace\nobreak 
    \noindent{\bf \the\secno. #1}\par\futurelet\next\sp@ce} 
\def\subsection#1{\subsubno=0\global\advance\subno by 1 
     \gdef\labeltype{\seclabel}%
     \ifssf\else\goodbreak\beforesubspace\fi 
     \global\ssffalse\nobreak 
     \noindent{\it \the\secno.\the\subno. #1\par}%
     \futurelet\next\subsp@ce} 
\def\subsubsection#1{\global\advance\subsubno by 1 
     \gdef\labeltype{\seclabel}%
     \ifssf\else\goodbreak\beforesubsubspace\fi 
     \global\ssffalse\nobreak 
     \noindent{\it \the\secno.\the\subno.\the\subsubno. #1}\null.  
     \ignorespaces} 
%
 
%
%
\def\numappendix#1{\ifappendix\ifnumbysec\countno=1\fi\else 
    \countno=1\figno=0\tabno=0\fi 
    \subno=0\global\advance\appno by 1 
    \secno=\appno\gdef\applett{A}\gdef\labeltype{\seclabel}%
    \global\appendixtrue\global\numapptrue 
    \goodbreak\beforesecspace\nobreak 
    \noindent{\bf Appendix \the\appno. #1\par}%
    \futurelet\next\sp@ce} 
\def\numsubappendix#1{\global\advance\subno by 1\subsubno=0 
    \gdef\labeltype{\seclabel}%
    \ifssf\else\goodbreak\beforesubspace\fi 
    \global\ssffalse\nobreak 
    \noindent{\it A\the\appno.\the\subno. #1\par}%
    \futurelet\next\subsp@ce} 
\def\@ppendix#1#2#3{\countno=1\subno=0\subsubno=0\secno=0\figno=0\tabno=0 
    \gdef\applett{#1}\gdef\labeltype{\seclabel}\global\appendixtrue 
    \goodbreak\beforesecspace\nobreak 
    \noindent{\bf Appendix#2#3\par}\futurelet\next\sp@ce} 
\def\Appendix#1{\@ppendix{A}{. }{#1}} 
\def\appendix#1#2{\@ppendix{#1}{ #1. }{#2}} 
\def\App#1{\@ppendix{A}{ }{#1}} 
\def\app{\@ppendix{A}{}{}} 
\def\subappendix#1#2{\global\advance\subno by 1\subsubno=0 
    \gdef\labeltype{\seclabel}%
    \ifssf\else\goodbreak\beforesubspace\fi 
    \global\ssffalse\nobreak 
    \noindent{\it #1\the\subno. #2\par}%
    \nobreak\subspace\noindent\ignorespaces} 
%
%
\def\@ck#1{\ifletter\bigskip\noindent\ignorespaces\else 
    \goodbreak\beforesecspace\nobreak 
    \noindent{\bf Acknowledgment#1\par}%
    \nobreak\secspace\noindent\ignorespaces\fi} 
\def\ack{\@ck{s}} 
\def\ackn{\@ck{}} 
\def\n@ip#1{\goodbreak\beforesecspace\nobreak 
    \noindent\smallfonts{\it #1}. \rm\ignorespaces} 
\def\naip{\n@ip{Note added in proof}} 
\def\na{\n@ip{Note added}} 
 
%
%
 
%
 
%
%
 
%
 
%
\def\table#1{\tablecaption{#1}} 
\def\tablecont{\topinsert\global\advance\tabno by -1 
    \tablecaption{(continued)}} 
\def\tablecaption#1{\gdef\labeltype{\tablabel}\global\widefalse 
    \leftskip=\secindent\parindent=0pt 
    \global\advance\tabno by 1 
    \smallfonts{\bf Table \ifappendix\applett\fi\the\tabno.} \rm #1\par 
    \smallskip\futurelet\next\t@b} 
\def\endtable{\vfill\goodbreak} 
\def\t@b{\ifx\next*\let\next=\widet@b 
             \else\ifx\next[\let\next=\fullwidet@b 
                      \else\let\next=\narrowt@b\fi\fi 
             \next} 
\def\widet@b#1{\global\widetrue\global\notfulltrue 
    \t@bwidth=\hsize\advance\t@bwidth by -\secindent}  
\def\fullwidet@b[#1]{\global\widetrue\global\notfullfalse 
    \leftskip=0pt\t@bwidth=\hsize}                   
\def\narrowt@b{\global\notfulltrue} 
\def\align{\catcode`?=13\ifnotfull\moveright\secindent\fi 
    \vbox\bgroup\halign\ifwide to \t@bwidth\fi 
    \bgroup\strut\tabskip=1.2pc plus1pc minus.5pc} 
\def\endalign{\egroup\egroup\catcode`?=12} 
 
%
%
\def\L#1{#1\hfill}

%
%
\def\br{\noalign{\vskip2pt\hrule height1pt\vskip2pt}} 
\def\mr{\noalign{\vskip2pt\hrule\vskip2pt}} 
\def\tabnote#1{\vskip-\lastskip\noindent #1\par} 
%
%

%
 
\catcode`?=13 
\def\lineup{\setbox0=\hbox{\smallfonts\rm 0}%
    \digitwidth=\wd0%
    \def?{\kern\digitwidth}%
    \def\\{\hbox{$\phantom{-}$}}%
    \def\-{\llap{$-$}}} 
\catcode`?=12 
%
%
\def\sidetable#1#2{\hbox{\ifppt\hsize=18pc\t@bwidth=18pc 
                          \else\hsize=15pc\t@bwidth=15pc\fi 
    \parindent=0pt\vtop{\null #1\par}%
    \ifppt\hskip1.2pc\else\hskip1pc\fi 
    \vtop{\null #2\par}}}  
\def\lstable#1#2{\everypar{}\tempval=\hsize\hsize=\vsize 
    \vsize=\tempval\hoffset=-3pc 
    \global\tabno=#1\gdef\labeltype{\tablabel}%
    \noindent\smallfonts{\bf Table \ifappendix\applett\fi 
    \the\tabno.} \rm #2\par 
    \smallskip\futurelet\next\t@b} 
\def\inctabno{\global\advance\tabno by 1} 
%
%
 
%
 
%
\def\figure#1{\figc@ption{#1}\bigskip} 
\def\figc@ption#1{\global\advance\figno by 1\gdef\labeltype{\figlabel}%
   {\parindent=\secindent\smallfonts\hang 
    {\bf Figure \ifappendix\applett\fi\the\figno.} \rm #1\par}} 
%
%
\def\refHEAD{\goodbreak\beforesecspace 
     \noindent\textfonts{\bf References}\par 
     \let\ref=\rf 
     \nobreak\smallfonts\rm} 
\def\references{\refHEAD\parindent=0pt 
     \everypar{\hangindent=18pt\hangafter=1 
     \frenchspacing\rm}%
     \secspace} 
\def\rf#1{\par\noindent\hbox to 21pt{\hss #1\quad}\ignorespaces} 
\def\refjl#1#2#3#4{\noindent #1 {\it #2 \bf #3} #4\par} 
\def\refbk#1#2#3{\noindent #1 {\it #2} #3\par} 
%
%
 
%
%
 
%
%

%
%

%
\catcode`\@=12 
%
%
 
%
%
\def\jnlstyle{\pptfalse\headsize{14}{18}%
\textsize{10}{12}%
\smallsize{8}{10} 
\textind=16pt} 
%
%
 
%
%
 
%
\parindent=\textind 
%
\input epsf
\def\received#1{\insertspace 
     \parindent=\secindent\ifppt\textfonts\else\smallfonts\fi 
     \hang{Received #1}\rm } 
\def\figure#1{\global\advance\figno by 1\gdef\labeltype{\figlabel}%
   {\parindent=\secindent\smallfonts\hang 
    {\bf Figure \ifappendix\applett\fi\the\figno.} \rm #1\par}} 
\headline={\ifodd\pageno{\ifnum\pageno=\firstpage\titlehead
   \else\rrhead\fi}\else\lrhead\fi} 
\def\lpsn#1#2{LPSN-#1-LT#2}
\def\endtable{\parindent=\textind\textfonts\rm\bigskip} 

\footline={\ifnum\pageno=\firstpage{\smallfonts cond-mat/9408077}
\hfil\textfonts\rm\folio\fi}   
\def\titlehead{\smallfonts J. Phys. A: Math. Gen. {\bf 27} (1994)
6349--6366   \hfil\lpsn{94}{3}} 
\firstpage=6349
\pageno=6349

\jnlstyle
\jl{1}
\overfullrule=0pt

\title{Surface magnetization of aperiodic Ising systems: a~comparative 
study of the bond and site problems}[Surface magnetization of
aperiodic Ising  systems]
 
\author{L Turban, P E Berche and 
B Berche}[L Turban \etal]
 
\address{Laboratoire de Physique du Solide\footnote\dag{Unit\'e de
Recherche Associ\'ee au CNRS No 155},  Universit\'e Henri Poincar\'e (Nancy~I),
BP 239, F--54506 Vand\oe uvre l\`es Nancy Cedex, France}

\received{27 May 1994}

\abs
We investigate the influence of aperiodic perturbations on the
critical behaviour at a second order phase transition. The bond and site
problems are compared for layered systems and aperiodic sequences generated
through substitution. In the bond problem, the interactions between the
layers are distributed according to an aperiodic sequence whereas in the site
problem, the layers themselves follow the sequence. A relevance-irrelevance
criterion introduced by Luck for the bond problem is extended to discuss the
site problem. It involves a wandering exponent for pairs, which can be larger
than the one considered before in the bond problem. The surface magnetization
of the layered two--dimensional Ising model is obtained, in the extreme
anisotropic limit, for the period--doubling and Thue--Morse sequences.
\endabs

\pacs{05.50.+q, 64.60.Cn, 64.60.Fr}
\submitted
\date

\section{Introduction} 
The discovery of quasicrystals (Shechtman \etal 1984) has opened a new
field of research which has been quite active during the last ten years 
(see Henley 1897, Janssen 1988, Janot \etal 1989, Guyot \etal
1991, Steinhardt and DiVicenzo 1991). On the  theoretical side
quasiperiodic or, more generally, aperiodic systems are interesting because
they appear as intermediates between periodic and random ones. Thus, phase
transitions in such systems are  expected to display a rich and unusual
critical behaviour.

Studies of the Ising model (Godreche \etal 1986, Okabe and
Niizeki 1988, S\o rensen \etal 1991), the percolation
problem (Sakamoto \etal 1989, Zhang and De'Bell 1993) and the statistics of
self--avoiding  walks (Langie and Igl\'oi 1992) on the two--dimensional
Penrose lattice did not show any change of critical exponents. Universal
behaviour was obtained in three dimensions too (Okabe and Niizeki 1990). But
some systems were also found for which the aperiodicity has some influence.
One may mention interface roughening in two dimensions for which a
continuously varying roughness exponent was obtained with the Fibonacci 
sequence (Henley and Lipowsky 1987, Garg and Levine 1987).

Some exact results have been obtained for the layered Ising model with an
aperiodic modulation of the interlayer
couplings (Igl\'oi 1988, Doria and
Satija 1988, Benza 1989, Henkel and Patk\'os 1992, Lin and Tao
1992, Turban and Berche 1993). The problem was studied in the extreme
anisotropic limit where it leads to a one--dimensional aperiodic quantum
Ising model (QIM) in a  transverse field, which is often
easier to handle (Kogut 1979). For the Fibonacci and other sequences, the
specific heat was found to display the Onsager logarithmic singularity but
for different sequences the singularity is washed out (Tracy 1988), like in
the randomly layered Ising model (McCoy and Wu 1968a, 1968b, McCoy 1970).

The situation was clarified in a recent study of the bulk properties of the
QIM (Luck 1993a). In this work, Luck proposed a generalization for
aperiodic systems of the Harris criterion (Harris 1974), allowing a
classification of critical aperiodic systems according to the sign of a
crossover exponent $\Phi$. This exponent involves the correlation length
exponent of the unperturbed system $\nu$ and the wandering exponent $\omega$
governing the size--dependence of the fluctuations of the aperiodic
interactions (Dumont 1990). The criterion has been also applied
to the case of anisotropic critical systems with uniaxial
aperiodicity (Igl\'oi 1993), explaining the interface roughening results.
It was later generalized to $d$--dimensional aperiodicities
in isotropic critical systems (Luck 1993b). The form of the singularities
with a relevant aperiodic perturbation has been discussed by
Igl\'oi using scaling arguments (Igl\'oi 1993). Recently, some exact
results for the surface magnetization of the QIM have also been obtained
for irrelevant, marginal and relevant
aperiodicities (Turban \etal 1994, Igl\'oi and Turban 1994) and conformal
aspects have been discussed (Grimm and Baake 1994). 

Most of the systems treated so far were dealing with an aperiodic
distribution of the couplings, i.e. with the {\it bond problem}. In
magnetic systems, this corresponds to an aperiodic distribution of the atoms
mediating the interactions in a superexchange mechanism. In the present
work, we study the surface magnetization of the aperiodic QIM, comparing 
the bond problem examined previously (Turban \etal 1994, Igl\'oi and
Turban 1994) to the {\it site problem} for which the magnetic moments are
distributed aperiodically and interact through a direct exchange mechanism.
Then the couplings depend on the nature of the two atoms involved in the
interaction. We show that, for a given aperiodic sequence, the 
perturbation may be more efficient for the site than for the bond problem
and may lead to a different critical behaviour. Exact results are obtained
for two typical aperiodic sequences.

In section~2 we present the Hamiltonian of the QIM and give 
the expression of the surface magnetization, defining the parameters for the
bond and site problems. In section~3 we recall the properties of
the substitution matrix, associated with an aperiodic sequence generated
through an inflation rule, for the bond problem. The substition matrix
adapted to the site problem is defined and compared to the previous one in
section~4. Then the relevance--irrelevance criterion is discussed
(section~5) and some general results about the QIM
critical coupling and surface magnetization are presented
(section~6). The period--doubling and Thue--Morse sequences are
studied in  sections~7 and 8 and the final section
contains a summary and discussion.

\section{Hamiltonian and surface magnetization}
Let us consider a layered semi--infinite two--dimensional Ising model with
exchange interactions $K_1(k)$ parallel to the surface and $K_2(k)$
between the layers at $k$ and $k\!+\!1$ (in $k_BT$ units). The extreme
anisotropic limit (Kogut 1979) corresponds to $K_1(k)\to\infty$,
$K_2(k)\to 0$ while keeping the ratio $\lambda_k=K_2(k)/K_1^*(k)$ fixed. In
this expression $K_1^*(k)\!=\!-1/2\ \ln\tanh K_1(k)$ is a dual coupling
which goes to zero in the limit. Introducing a constant reference coupling
$K_1^*$, the dual coupling can be rewritten as $h_k K_1^*$ where $h_k$
is the transverse field. The transfer operator
$\exp(-2K_1^*{\cal H})$ involves the Hamiltonian of a one--dimensional spin
$1/2$ quantum chain. Introducing the two--spin interactions 
$J_k=h_k\lambda_k$, the QIM takes the following form 
$$
{\cal H}=-{1\over 2}\sum_{k=1}^\infty
[h_k\sigma_k^z+J_k\sigma_k^x\sigma_{k+1}^x]
\eqno(2.1)
$$
where the $\sigma$s are Pauli spin operators.

The Hamiltonian can be put in diagonal form (Lieb \etal 1961) 
$$
{\cal H}=\sum_\nu\epsilon_\nu\left(\eta_\nu^\dagger\eta_\nu-{1\over
2}\right) 
\eqno(2.2)
$$
using the Jordan--Wigner transformation (Jordan and Wigner 1928) followed by
a canonical transformation to the diagonal fermion operators $\eta_\nu$. 
The fermion excitation spectrum is obtained as the solution of the
eigenvalue problem
$$
\eqalign{
&\epsilon_\alpha\psi_\alpha(k)=-h_k\phi_\alpha(k)-J_k 
\phi_\alpha(k+1)\cr	
&\epsilon_\alpha\phi_\alpha(k)=-J_{k-1}\psi_\alpha(k-1)-h_k
\psi_\alpha(k)\cr
&h_0=J_0=0\cr}\eqno(2.3)
$$
where the $\phi_\alpha(k)$ and $\psi_\alpha(k)$ are the components of
two normalized eigenvectors which satisfy the boundary conditions
$\phi_\alpha(0)\!=\!\psi_\alpha(0)\!=\! 0$. 

In the ordered phase, the two--spin correlation function in the surface
asymptotically gives the square of the surface magnetization, which can be
written as the matrix element $m_s=<\! 1\vert\sigma_1^x\vert 0\!>$ between
the ground state and the first excited state of the Hamiltonian. For the
semi--infinite system these two states become degenerate in the ordered 
phase, i.e. the lowest excitation $\epsilon_1$ vanishes. Using the
above--mentioned transformation to diagonal fermions, it can be shown that
$m_s$ is also given by $\phi_1(1)$. According to the first equation
in~(2.3) with $\epsilon_\alpha=0$, the other components of the
eigenvector can be deduced from the recursion relation
$$
\phi_{1}(k+1)=-{h_k\over J_k}\phi_1(k)=-\lambda_k^{-1}\phi_1(k).
\eqno(2.4)
$$
The normalization of the
eigenvector then leads to the surface magnetization (Peschel~1984)
$$
m_s=\left(1+\sum_{j=1}^\infty\prod_{k=1}^j\lambda_k^{-2}\right)^{-1/2},
\eqno(2.5) 
$$
where the couplings $J_k$ and $h_k$ in ${\cal H}$ only enter through their
ratio $\lambda_k$. Thus $m_s$ is the same as for a quantum
chain with $h_k=1$ and an effective two--spin interaction $\lambda_k$.
Such a reparametrization of the Hamiltonian is no longer possible when
nonvanishing excitations are involved, if the transverse field is
$k$ dependent. It follows that, in general, the effect of both interactions
have to be considered. We shall come back to this point
in section~5.

In the bond problem the interactions parallel to the surface are
constant, $K_1(k)\!=\! K_1$, so that $h_k\!=\! 1$ and
$\lambda_k\!=\! K_2(k)/K_1^*$  depends on the layer index $k$ only through
the value of the interlayer interaction. In the site problem, the effective
coupling $\lambda_k\!=\! K_2(k)/K_1^*(k)$ involves both the
interaction inside layer $k$ and the interaction between
layers $k$ and $k\!+\! 1$. As a consequence, its value depends
on the nature of the two layers and it is generally asymmetric. 
 
\section{Substitution matrix for the bond problem}
In this section we give a brief summary of the properties of
aperiodic sequences generated through an inflation rule. To simplify the
presentation, we consider sequences involving only two letters $A$ and $B$.
In the bond problem these letters correspond to the interlayer interactions
$\lambda_A$, $\lambda_B$. The generalization to any
number of letters is straightforward. 

A sequence is constructed through iterated substitutions on the two
letters, $A\to{\cal S}\{ A\}$, $B\to{\cal S}\{ B\}$. The process will be
illustrated on the following example:
$$
\eqalign{
{\cal S}\{ A\}&=B\  A\  A,\cr 
{\cal S}\{ B\}&=A\  B.\cr}
\eqno(3.1)
$$
When the construction starts on $A$, after $n$ steps, one obtains:
$$
\eqalign{
&n=0\qquad A\cr
&n=1\qquad B\  A\  A\cr
&n=2\qquad A\  B\  B\  A\  A\  B\  A\  A\cr
&\dots\cr}
\eqno(3.2)
$$
whereas, starting on $B$, the iteration gives
$$
\eqalign{
&n=0\qquad B\cr
&n=1\qquad A\  B\cr
&n=2\qquad B\  A\  A\  A\  B\cr
&\dots\cr}
\eqno(3.3)
$$
Informations about the sequence
are contained in the substitution matrix   
$$
{\bss M_1}=\left(\matrix{n_A^{{\cal S}\{ A\}}&n_A^{{\cal S}\{ B\}}\cr
                      n_B^{{\cal S}\{ A\}}&n_B^{{\cal S}\{ B\}}\cr}
\right)=\left(\matrix{2&1\cr 1&1\cr}\right) ,
\eqno(3.4)
$$
where the matrix elements give the numbers of $A$ or $B$ in ${\cal S}\{
A\}$ or ${\cal S}\{ B\}$. It is easy to check that the numbers $L_n^A$
and $L_n^B$ of $A$ and $B$ in the sequence, after $n$
substitutions, are given by the corresponding matrix elements in
${\bss M_1}^n$. They belong to the first (second) column when the
construction starts on $A$ $(B)$. 

Let $\bi{V}\!_\alpha$ be the right eigenvectors
and $\Lambda_\alpha$ the eigenvalues of the substitution matrix such that 
$$
{\bss M_1}\bi{V}\!_\alpha=\Lambda_\alpha
\bi{V}\!_\alpha.
\eqno(3.5)
$$
$L_n^A$, $L_n^B$ and the length of the sequence, $L_n$, are asymptotically
proportional to $\Lambda_1^n$ where $\Lambda_1>1$ is the eigenvalue  of 
${\bss M_1}$ with largest modulus, which is real and positive
according to the Perron--Frobenius theorem. The asymptotic density,
$\rho_\infty^A=\lim_{n\to\infty}L_n^A/L_n$, can be deduced from the
associated eigenvector with 
$$
\rho_\infty^A=1-\rho_\infty^B={V_1(1)\over V_1(1)+V_1(2)}.
\eqno(3.6)
$$

The interactions in the bond problem can be rewritten as 
$$
\lambda_A=\overline{\lambda}+\rho_\infty^B\delta_1\  ,\quad
\lambda_B=\overline{\lambda}-\rho_\infty^A\delta_1\  , 
\eqno(3.7)
$$
where $\overline{\lambda}$ is the averaged coupling and
$\delta_1\!=\!\lambda_B\!-\!\lambda_A$, the amplitude of the aperiodic
modulation. At a length scale $L_n$, the cumulated deviation from the
average is  
$$
\Delta_1(L_n)=\sum_{k=1}^{L_n}\left(\lambda_k-\overline{\lambda}\right)
\sim\delta_1\Lambda_2^n\sim\delta_1 L_n^{\omega_1}
\eqno(3.8)
$$
where $\Lambda_2$ is the second largest eigenvalue in modulus. The
wandering exponent   
$$
\omega_1={\ln\vert\Lambda_2\vert\over\ln\Lambda_1}
\eqno(3.9)
$$
governs the behaviour of the fluctuations of the interlayer couplings 
(Dumont 1990).

\section{Substitution matrix for the site problem} 
Let us now consider the site problem. The two letters $A$ and $B$ then
correspond to the two magnetic species which are distributed according to
the aperiodic sequence, with the layers containing either $A$ or $B$
atoms. The QIM effective couplings $\lambda_{AA}$,
$\lambda_{AB}$, $\lambda_{BA}$ and $\lambda_{BB}$ depend on the nature of
the two layers involved in the interaction through the intra-- and
interlayer couplings in the anisotropic classical system. 

In order to count the numbers of bonds of different types in the sequence
after $n$ substitutions, $L_n^{AA}$, $L_n^{AB}$, $L_n^{BA}$ and $L_n^{BB}$,
one defines the two--letter substitution matrix 
\medskip
$$
\fl{\bss M_2}=\left(\matrix{
n_{AA}^{{\cal S}\{ A[A]\}}&n_{AA}^{{\cal S}\{ A[B]\}}&
n_{AA}^{{\cal S}\{ B[A]\}}&n_{AA}^{{\cal S}\{ B[B]\}}\cr
n_{AB}^{{\cal S}\{ A[A]\}}&n_{AB}^{{\cal S}\{ A[B]\}}&
n_{AB}^{{\cal S}\{ B[A]\}}&n_{AB}^{{\cal S}\{ B[B]\}}\cr
n_{BA}^{{\cal S}\{ A[A]\}}&n_{BA}^{{\cal S}\{ A[B]\}}&
n_{BA}^{{\cal S}\{ B[A]\}}&n_{BA}^{{\cal S}\{ B[B]\}}\cr
n_{BB}^{{\cal S}\{ A[A]\}}&n_{BB}^{{\cal S}\{ A[B]\}}&
n_{BB}^{{\cal S}\{ B[A]\}}&n_{BB}^{{\cal S}\{ B[B]\}}\cr}
\right)
=\left(\matrix{1&2&0&0\cr 1&0&1&1\cr 
                        1&1&0&1\cr 0&0&1&0\cr}\right) ,
\eqno(4.1)
$$
where for example, $n_{AB}^{{\cal S}\{ B[A]\}}$ gives the number of
$AB$--bonds in the sequence generated by ${\cal S}\{ B\}$ complemented by
the first letter in ${\cal S}\{ A\}$. Such a sequence builds the first part
of the sequence which is obtained when the inflation rule is applied to a
$BA$--bond. The same matrix ${\bss M_2}$ has been
considered before as resulting from substitutions on words of length two
(Queff\'elec 1987). 

With the example of the previous section, the two--letter substitutions
read   
$$ \eqalign{
{\cal S}\{ A[A]\}&=B\  A\  A\  [B]\cr
{\cal S}\{ A[B]\}&=B\  A\  A\  [A]\cr
{\cal S}\{ B[A]\}&=A\  B\  [B]\cr
{\cal S}\{ B[B]\}&=A\  B\  [A]\cr}
\eqno(4.2)
$$
and lead to the last matrix in equation~(4.1).

As before, the matrix elements in each column of ${\bss M_2}^n$
give the numbers of bonds of each type ($L_n^{ij}$; $i,j=A,B$) in the
sequence obtained after $n$ iterations. These numbers are found in the two
first (last) columns if the construction starts on $A$ ($B$) and are
given by the minimum of the two values appearing in each half--row.
Due to end effects, one of the values in each column differs by $1$ from
the true number of bonds in the sequence. At $n=2$, for example, we have  
$$
{\bss M_2}^2=\left(\matrix{3&2&2&2\cr 2&3&1&1\cr 
                          2&2&2&1\cr 1&1&0&1\cr}\right). 
\eqno(4.3)
$$
The first two columns correspond to a sequence constructed on $A$ at the
second iteration and lead to $L_2^{AA}\!=\!2$, $L_2^{AB}\!=\!2$,
$L_2^{BA}\!=\!2$ and $L_2^{BB}\!=\!1$, in agreement with~(3.2).

Since the sum of the numbers of bonds 
starting with a given letter (i.e. with $A$, $AA$ and
$AB$ bonds) gives the number of times this letter is met in the sequence for
the bond problem, the matrix elements of ${\bss M_1}$ are recovered by
taking the sum of the two first elements and the sum of the two last
elements in each column of ${\bss M_2}$. The same results are obtained
with the two first (last) columns since the corresponding sequences only
differ through their last bond. The same relation exists between the
elements of ${\bss M_2}^n$ and ${\bss M_1}^n$.

Let $\Omega_\alpha$ be the eigenvalues and $\bi{W}\!_\alpha$ the right
eigenvectors of ${\bss M_2}$. The numbers of bonds $L_n^{ij}$
($i,j\!=\!A,B$) in the sequence after $n$ iterations are still
proportional to the $n$th power of the largest eigenvalue $\Omega_1$.
Using the associated eigenvector, one obtains the asymptotic bond densities
$$
\eqalign{
\rho_\infty^{AA}&={W_1(1)\over\sum_{i=1}^4W_1(i)}\quad  
\rho_\infty^{AB}={W_1(2)\over\sum_{i=1}^4W_1(i)}\cr 
\rho_\infty^{BA}&={W_1(3)\over\sum_{i=1}^4W_1(i)}\quad  
\rho_\infty^{BB}={W_1(4)\over\sum_{i=1}^4W_1(i)} .\cr} 
\eqno(4.4)
$$

Since the length of the sequence after $n$ steps is also the sum of the
$L_n^{ij}$, the leading eigenvalues of the two matrices are the same.
Using the above--mentioned relation between the matrix elements of
${\bss M_2}$ and ${\bss M_1}$, the secular equation of
${\bss M_2}$  can be factorized. The first factor gives back the
secular equation of ${\bss M_1}$ so that $\Omega_1=\Lambda_1$ and
$\Lambda_2$ also belongs to the spectrum of ${\bss M_2}$ . The two
last  eigenvalues of ${\bss M_2}$ follow from the second factor and read
$$  
\Omega_\alpha={1\over 2}\left[ a+b\pm\sqrt{(a-b)^2+4cd}\right]
\eqno(4.5)
$$
where
$$
\eqalign{
a&=n_{AA}^{{\cal S}\{A[A]\}}-n_{AA}^{{\cal S}\{A[B]\}}\qquad
b=n_{BA}^{{\cal S}\{B[A]\}}-n_{BA}^{{\cal S}\{B[B]\}}\cr
c&=n_{BA}^{{\cal S}\{A[A]\}}-n_{BA}^{{\cal S}\{A[B]\}}\qquad
d=n_{AA}^{{\cal S}\{B[A]\}}-n_{AA}^{{\cal S}\{B[B]\}}.\cr}
\eqno(4.6)
$$
Since the coefficients in~(4.6) involve differences between the
numbers of bonds in sequences which at most differ through their last
bond, they are equal to $0$ or $\pm 1$. They are completely
determined through the first and last letters in ${\cal S}\{A\}$ and ${\cal
S}\{ B\}$ and can be obtained by inspection. The two eigenvalues are
also equal to $0$ or $\pm 1$. When the two substitutions begin with the same
letter, the coefficients and the eigenvalues $\Omega_\alpha$ vanish. Other
cases are listed in table~1. 
\bigskip
\table{Coefficients of the secular equation~(4.5) and corresponding
eigenvalues for substitutions starting with different letters.}[w]
\lineup
\align\L{#}&\L{#}&\L{#}&\L{#}&\L{#}&\L{#}&\L{#}&\L{#}&\L{#}\cr
\br
${\cal S}\{A\}$&$A\dots A$&$A\dots A$&$A\dots B$&$A\dots B$&$B\dots A$
&$B\dots A$&$B\dots B$&$B\dots B$\cr
${\cal S}\{B\}$&$B\dots A$&$B\dots B$&$B\dots A$&$B\dots B$&$A\dots A$
&$A\dots B$&$A\dots A$&$A\dots B$\cr
\mr
$a$&$1$&$1$&$0$&$0$&$\lo-1$&$\lo-1$&$0$&$0$\cr
$b$&$0$&$1$&$0$&$1$&$0$&$\lo-1$&$0$&$\lo-1$\cr
$c$&$1$&$0$&$1$&$0$&$\lo-1$&$0$&$\lo-1$&$0$\cr
$d$&$0$&$0$&$1$&$1$&$0$&$0$&$\lo-1$&$\lo-1$\cr
$\Omega_\alpha$&$1$, $0$&$1$, $1$&$1$, $-1$&$1$, $0$&$\lo-1$,
$0$&$\lo-1$, $-1$&$1$, $-1$ &$\lo-1$, $0$\cr 
\br
\endalign
\endtable

Like in the bond problem, the fluctuations in the couplings $\lambda_k$
can be deduced from the substitution matrix, working in the basis of the
right eigenvectors $\bi{W}\!_\alpha$. The averaged coupling
$\overline{\lambda}$ is now given by 
$$
\overline{\lambda}=\rho_\infty^{AA}\lambda_{AA}+\rho_\infty^{AB}
\lambda_{AB}
+\rho_\infty^{BA}\lambda_{BA}+\rho_\infty^{BB}\lambda_{BB}.
\eqno(4.7)
$$
The cumulated deviation from $\overline{\lambda}$ is generally governed by
$\Omega_2$, the second largest eigenvalue of ${\bss M_2}$ in modulus 
with, at a length scale $L_n$, 
$$
\Delta_2(L_n)=\sum_{k=1}^{L_n}\left(\lambda_k-\overline{\lambda}\right)
\sim\delta_2\Omega_2^n\sim\delta_2 L_n^{\omega_2}.
\eqno(4.8)
$$
Here, the amplitude of the perturbation $\delta_2$ is a linear combination
of coupling differences, $\lambda_{AA}-\lambda_{AB}$,
$\lambda_{AA}-\lambda_{BA}$, $\dots$, and the wandering exponent for
bonds can be written as 
$$
\omega_2={\ln\vert\Omega_2\vert\over\ln\Lambda_1}. 
\eqno(4.9)
$$

As a consequence, the wandering exponent  changes when
$\vert\Omega_2\vert\!=1 \!>\!\vert\Lambda_2\vert$. Some examples are
listed in the next section.  

\section{Relevance--irrelevance criterion}
In both problems the aperiodic modulation introduces a thermal
perturbation above $\overline{\lambda}$ which, at a length scale $L$, has
an averaged density
$$
\overline{\delta\lambda}_i(L)={\Delta_i(L)\over
L}\sim\delta_iL^{\omega_i-1}\qquad (i=1,2)
\eqno(5.1)
$$
where the $\omega_i$s are the wandering exponents defined in
equation~(3.9) for the bond problem and equation~(4.9) for the
site problem. 

Changing the length scale by $b\!=\! L/L'$ leads to the
renormalized density 
$$
\overline{\delta\lambda'}_i(L')\sim\delta_i'\left(L\over
b\right)^{\omega_i-1}\!\! =b^{1/\nu}\delta_i L^{\omega_i-1}\qquad (i=1,2),
\eqno(5.2)
$$
where $\nu$ is the bulk correlation length exponent. As a consequence,
the perturbation amplitude $\delta_i$ transforms like
$$
\delta_i'=b^{\omega_i-1+1/\nu}\delta_i
\eqno(5.3)
$$
with a crossover exponent 
$\Phi_i=1+\nu(\omega_i-1)$ (Luck 1993a, Igl\'oi 1993).

When $\Phi_i$ is positive, the perturbation grows under rescaling,
leading to a new fixed point with a different critical behaviour. When
$\Phi_i$ is negative,  the perturbation decays under rescaling and the
critical behaviour is the same as for the homogeneous system. The crossover
exponent vanishes in the marginal case and then the critical exponents may
vary continuously with the perturbation amplitude, i.e. the system may
display a nonuniversal critical behaviour.

For the site problem, as mentioned at the end of section~2, the
aperiodic modulation does not generally combine into a single effective
parameter and one has to consider its effect on the transverse field and 
the two--spin interaction separately. The transverse field may take the
values $h_A$ or $h_B$ like $\lambda_k$ in the bond problem. Thus, the
behaviour of the corresponding perturbation under rescaling is controlled
by $\Phi_1$, whereas the perturbation of the two--spin interaction $J_k$ is
governed by $\Phi_2$. Since, according to the discussion of
section~4, $\omega_2\!\geq\!\omega_1$, for any singular quantity
the critical behaviour will depend on the sign of $\Phi_2$ in the site
problem, like for the surface magnetization. 
\bigskip
\table{Comparison between the bond and site problems for some typical 
sequences. The last two lines refer to the $2d$ Ising model.}[w]
\lineup
\align\L{#}&\L{#}&\L{#}&\L{#}&\L{#}&\L{#}\cr
\br
sequence&Fibonacci$^a$&Thue--Morse$^b$
&period--&three--&fivefold$^e$\cr
&&&doubling$^c$&folding$^d$\cr
\mr
${\cal S}\{A\}$&$B$&$AB$&$BB$
&$ABA$&$AAAB$\cr
${\cal S}\{B\}$&$BA$&$BA$&$BA$
&$ABB$&$BBA$\cr
$\Lambda_1$&$\tau^f$&$2$
&$2$&$3$&$2+\tau$\cr
$\vert\Lambda_2\vert$&$\tau-1$&$0$
&$1$&$1$&$3-\tau$\cr
$\vert\Omega_2\vert$&$\tau-1$&$1$
&$1$&$1$&$3-\tau$\cr
$\omega_1$&$\lo-1$&$\lo-\infty $&$0$
&$0$&$0$.$25157$\cr
$\omega_2$&$\lo-1$&$0$&$0$
&$0$&$0$.$25157$\cr
bond problem&irrelevant&irrelevant
&marginal&marginal&relevant\cr
site problem&irrelevant&marginal
&marginal&marginal&relevant\cr
\br
\endalign
\tabnote{$^a$ See e.g. (Tracy 1988).}
\tabnote{$^b$ See e.g. (Dekking \etal 1983a).}
\tabnote{$^c$ This sequence appears in connection with the
period--doubling cascade  (Collet and Eckmann 1980).}
\tabnote{$^d$ This is the folding sequence of a dragon curve
(Dekking \etal 1983b).} 
\tabnote{$^e$ This sequence is connected with tilings  of the
plane with fivefold symmetry (Godr\'eche and Luck 1992, Godr\'eche and
Lan\c con 1992).} 
\tabnote{$^f$ $\tau\!=\!(1\!+\!\sqrt{5})/2$ is the golden
mean.}
\endtable

Table~2 gives the two largest eigenvalues of the substitution
matrices and the wandering exponents for typical aperiodic sequences. For
the two--dimensional Ising model with $\nu\!=\! 1$, the borderline  between
relevant and irrelevant behaviour corresponds to $\omega_i\!=\!0$
according to~(5.3). The bond and site aperiodic perturbations
are irrelevant for the Fibonacci sequence whereas the site
perturbation becomes marginal for the Thue--Morse sequence. The following
sequences all have divergent fluctuations so that the perturbation behaves
in the same way for both problems. The period--doubling and three--folding
sequences lead to marginal perturbations but the nonuniversal
exponents should differ for the bond and site problems. The last sequence
is relevant in both cases. 

\section{General results: critical coupling and surface magnetization}
The critical coupling of the inhomogeneous QIM is such that (Pfeuty 1979) 
$$
\lim_{L\to\infty}\prod_{k=1}^L\left({J_k\over h_k}\right)_c^{1/L}= 
\lim_{L\to\infty}\prod_{k=1}^L\left(\lambda_{k}\right)_c^{1/L}=1 .
\eqno(6.1)
$$
For the bond problem, we take $\lambda_A\!=\!\lambda$ as a reference
coupling and $\lambda_B\!=\!\lambda r$. We associate a digit $f_k$ with the
$k$th letter in the sequence. With $f_k\!=\! 0$ for $A$
and $f_k\!=\! 1$ for $B$, the $k$th coupling can be written as
$\lambda_k\!=\!\lambda r^{f_k}$. The critical coupling $\lambda_c$ is such
that $\lim_{L\to\infty}\lambda_cr^{n_L/L}=1$ with
$$
n_j=\sum_{k=1}^jf_k .
\eqno(6.2)
$$
Finally, one obtains
$$
\lambda_c=r^{-\rho_\infty}
\eqno(6.3)
$$
where $\rho_\infty=\rho_\infty^B$ defined in equation~(3.6).

For the site problem, we take the following parametrization:
$$
\lambda_{AA}=\lambda,\quad\lambda_{AB}=\lambda u,\quad
\lambda_{BA}=\lambda v,\quad\lambda_{BB}=\lambda r .
\eqno(6.4)
$$
Due to the transverse field contribution, $\lambda_{AB}$ is
generally different from $\lambda_{BA}$. The effective coupling can be
written as
$$
\lambda_k=\lambda r^{f_kf_{k+1}}u^{f_{k+1}-f_kf_{k+1}}v^{f_k-f_kf_{k+1}}
\eqno(6.5)
$$
and using (6.1), the critical coupling is such that
$$
\lim_{L\to\infty}u^{(f_{L+1}-f_1)/L}\lambda_cs^{n_L/L}\left({r\over
s}\right)^{m_L/L}=1
\eqno(6.6)
$$
where $s=uv$ and
$$
m_j=\sum_{k=1}^jf_kf_{k+1} .
\eqno(6.7)
$$
The critical coupling is then given by
$$
\lambda_c=s^{-\rho_\infty}\left({s\over r}\right)^{\kappa_\infty},\quad
\kappa_\infty=\lim_{L\to\infty}{m_L\over L}.
\eqno(6.8)
$$
Alternatively, using the asymptotic
densities defined in~(4.4), the critical coupling can be
expressed as   
$$
\lambda_c=u^{-\rho_\infty^{AB}}v^{-\rho_\infty^{BA}}r^{-\rho_\infty^{BB}} .
\eqno(6.9)
$$

According to equation~(2.5) and making use of~(6.2) with $n_0\!=\! 0$,
the surface magnetization for the bond problem is given by
$$
m_s=\left[S(\lambda,r)\right]^{-1/2},\quad 
S(\lambda,r)=\sum_{j=0}^\infty\lambda^{-2j}r^{-2n_j}.
\eqno(6.10)
$$
For the site problem, using equations~(2.5), (6.2),
(6.5), and (6.7) with $m_0=0$, we have 
$$
\eqalign{
m_s&=\left[\Sigma(\lambda,r,s,u)\right]^{-1/2},\cr
\Sigma(\lambda,r,s,u)&=u^{2f_1}\sum_{j=0}^\infty\lambda^{-2j}
\left({r\over s}\right)^{-2m_j}s^{-2n_j}u^{-2f_{j+1}}.\cr}
\eqno(6.11)
$$
Since $f_k=0,1$, one may use the identity 
$$
a^{f_k}=1+(a-1)f_k
\eqno(6.12)
$$ 
to rewrite the sum as
$$
\Sigma(\lambda,r,s,u)=u^{2f_1}\left[\Sigma_1\left(\lambda,{r\over
s},s\right) +(u^{-2}-1)\Sigma_2\left(\lambda,{r\over
s},s\right)\right],
\eqno(6.13)
$$
where 
$$
\eqalign{
\Sigma_1(\lambda,x,y)&=\sum_{j=0}^\infty\lambda^{-2j}x^{-2m_j}
y^{-2n_j},\cr
\Sigma_2(\lambda,x,y)&=\sum_{j=0}^\infty\lambda^{-2j}x^{-2m_j}
y^{-2n_j}f_{j+1}.\cr} 
\eqno(6.14)
$$

Let us consider the case $s\!=\!r$, i.e.
$\lambda_{AB}\lambda_{BA}\!=\!\lambda_{AA}\lambda_{BB}$, a common
approximation in the case of symmetric couplings
($\lambda_{AB}\!=\!\lambda_{BA}$). Then $\Sigma_1(\lambda,1,r)\!=\!
S(\lambda,r)$ and, using the identity
$f_{j+1}\!=\!(r^{-2f_{j+1}}-1)/(r^{-2}-1)$, one obtains
$$
\fl\Sigma(\lambda,r,r,u)=u^{2(f_1-1)}{1-u^2\over
1-r^2}\left[\left(\lambda^2r^2+{u^2-r^2\over
1-u^2}\right)S(\lambda,r)-\lambda^2r^2\right] . 
\eqno(6.15)
$$
It follows that, for this particular combination of couplings, the
critical behaviour is governed by $S(\lambda,r)$ and is the same as in the
bond problem. 

\section{Period--doubling sequence}
The results of the preceding sections will be now illustrated on the
examples of two sequences with different surface magnetization exponents
for the bond and site problems. We begin with the period--doubling
sequence (Luck 1993a, Collet and Eckmann 1980) which is marginal for both
problems, according to table~2. 

Since the solution of the bond problem can be
found elsewhere (Turban \etal 1994), we only give a summary of the results.
Starting the iteration on $B$ and using the substitutions given
in~table~2, one obtains   
$$
\eqalign{
&B\cr
&B\  A\cr
&B\  A\  B\  B\cr
&B\  A\  B\  B\  B\  A\  B\  A\cr
&B\ \underline{A}\  B\ \underline{B}\  B\ \underline{A}\  B\ 
 \underline{A}\  B\ \underline{A}\  B\ \underline{B}\  B\ 
 \underline{A}\  B\ \underline{B}\cr
&\dots\cr}
\eqno(7.1)
$$
The asymptotic density $\rho_\infty\!=\!2/3$ leads to
the critical coupling 
$$
\lambda_c=r^{-2/3}\qquad {\rm (bond\  problem)} .
\eqno(7.2)
$$
The form of the substitution is such that
$$
f_{2k}=1-f_k,\qquad f_{2k+1}=1, 
\eqno(7.3)
$$
from which one deduces
$$
n_{2j}=2j-n_j,\quad n_{2j+1}=2j+1-n_j .
\eqno(7.4)
$$
Splitting the sum $S(\lambda,r)$ in~equation~(6.10) into even and odd
parts and using (7.4), one obtains the recursion relation 
$$
S(\lambda,r)=\left(1+{1\over\lambda^2r^2}\right)S(\lambda^2r^2,r^{-1})
\eqno(7.5)
$$
and the series can be written as an infinite product (Turban \etal 1994)
$$
S(\lambda,r)=\prod_{k=1}^\infty\left[1+\lambda_c
{\left(\lambda_c\over\lambda\right)}^{2^{2k-1}}\right]\left[1+\lambda_c^{-1}
{\left(\lambda_c\over\lambda\right)}^{2^{2k}}\right] .
\eqno(7.6)
$$
Let $S(z)$ be the series expansion of
$S(\lambda,r)$ in powers of $z\!=\!(\lambda_c/\lambda)^2$. According to
equation~(6.10), near the critical point $S(z)$ should display a power
law singularity with $S(z)\!\sim\!(1-z)^{-2\beta_s}$, where $\beta_s$ is the
surface magnetization exponent. It may be shown that, at the
critical point $z\!=\!1$, the truncated series $S_L(z)$ containing the 
first $L$ terms in $S(z)$ behaves as $L^{2\beta_s}$ (Igl\'oi 1986). Since
the first $l$ terms in~(7.6) just contain the truncated series with
$L\!=\!2^{2l}$, the surface magnetization exponent is given
by (Turban \etal 1994)
$$
\beta_s={\ln\left[(1+\lambda_c)(1+\lambda_c^{-1})\right]\over 4\ln 2}
={1\over 2}{\ln(\lambda_c^{1/2}+\lambda_c^{-1/2})\over\ln 2}.
\eqno(7.7)
$$

{\par\begingroup\parindent=0pt\medskip
\epsfxsize=9truecm
\topinsert
\centerline{\epsfbox{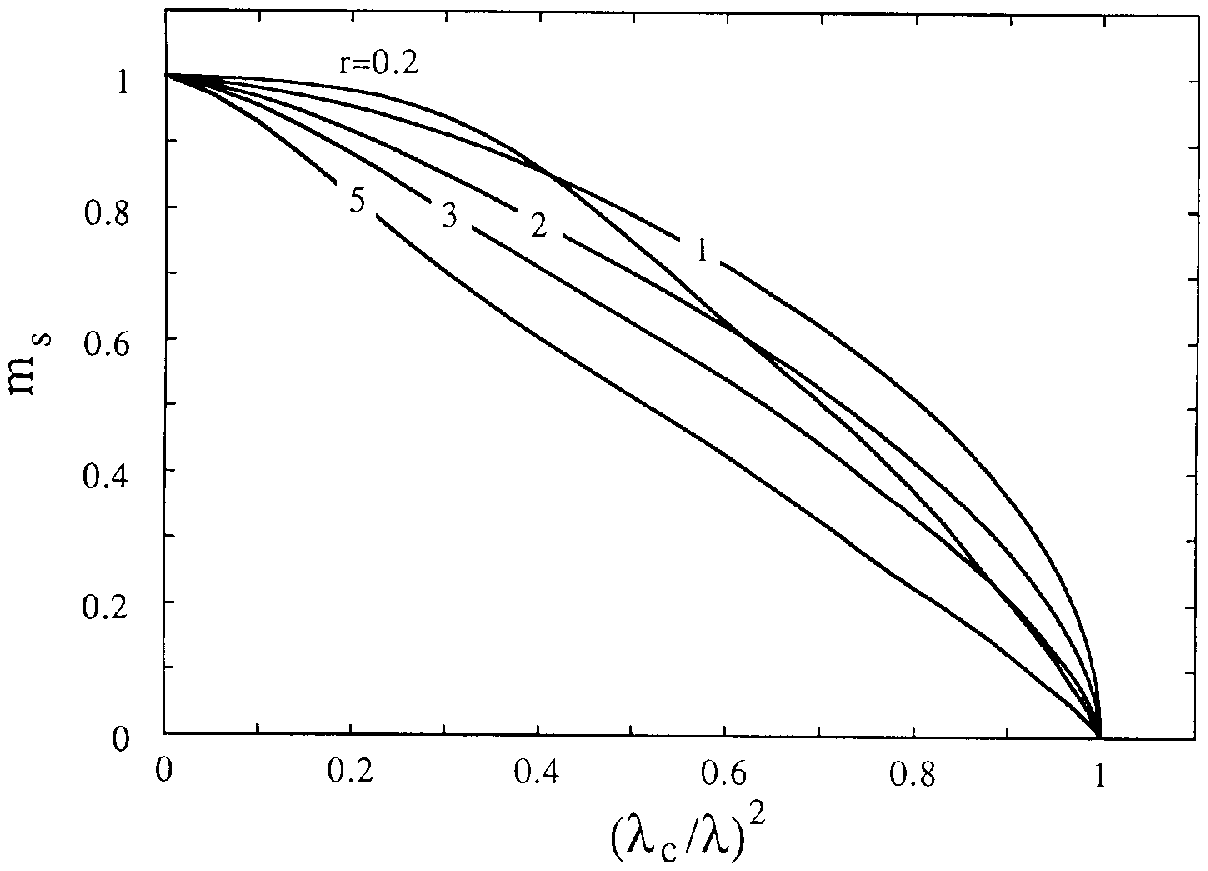}}
\smallskip
\figure{Temperature dependence of the surface magnetization
(period--doubling, site problem) for different values of $r$,
$u\!=\!.5$ and $v\!=\!2$.}
\endinsert
\endgroup
\par}

{\par\begingroup\medskip
\epsfxsize=9truecm
\midinsert
\centerline{\epsfbox{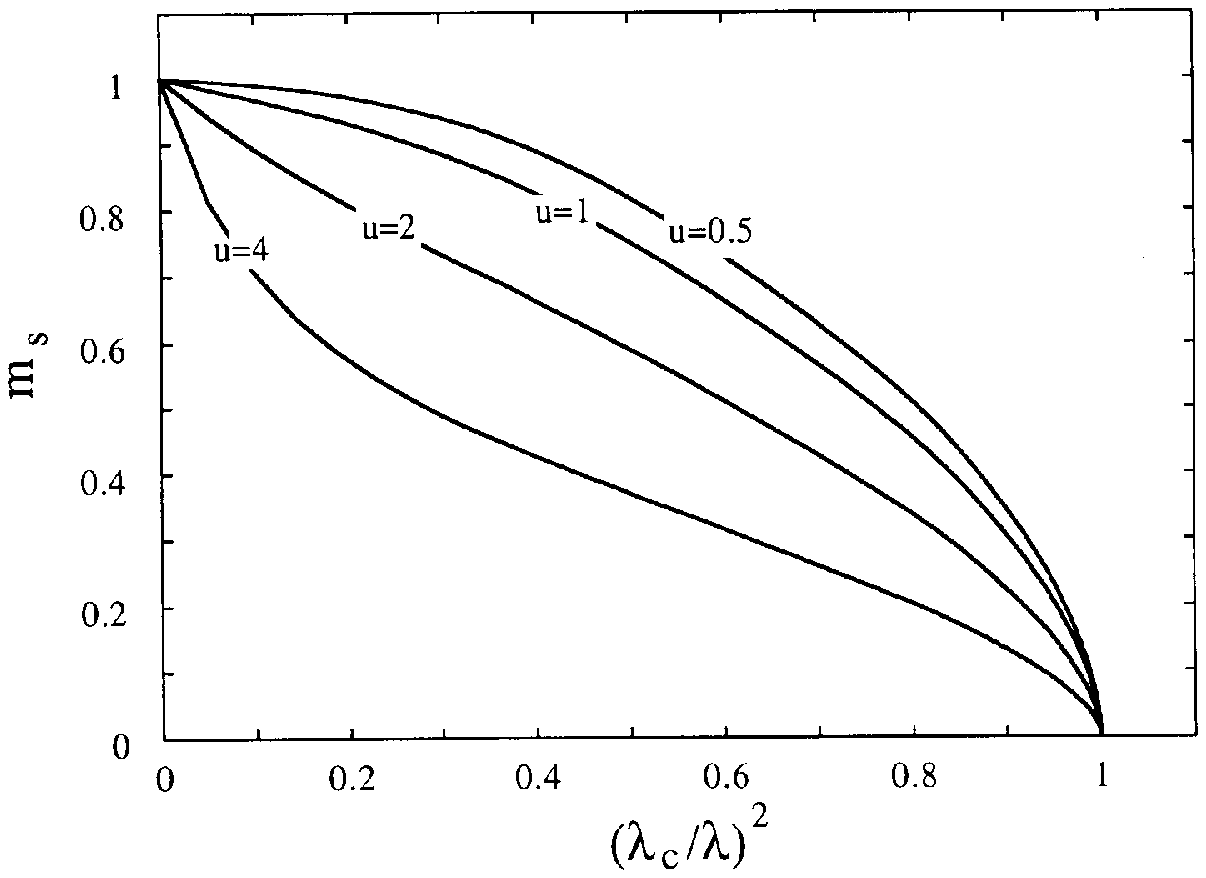}}
\smallskip
\figure{Temperature dependence of the surface
magnetization (period--doubling, site problem) for different values of
$u$, $r\!=\!.5$ and $s\!=\!1$.}
\endinsert
\endgroup
\par}

Let us now consider the site problem. The critical
coupling, 
$$
\lambda_c=(rs)^{-1/3}\qquad {\rm(site\  problem}),
\eqno(7.8)
$$
follows from~(4.4) and (6.9)
with $\rho_{AB}\!=\!\rho_{BA}\!=\!\rho_{BB}\!=\! 1/3$. Putting (7.3)
into (6.7), one obtains the recursion relations
$$
m_{2j}=2j-2n_j,\quad m_{2j+1}=2j+1-2n_j-f_{j+1},
\eqno(7.9)
$$
which can be used, together with those given in~(7.4), to
relate $\Sigma_1$, $\Sigma_2$ and $S$. Splitting, as above, the sums in
equation~(6.14) into odd and even values of $j$, and using the identity
$$
a^{f_k}={b-a\over b-1}+{a-1\over b-1}f_k,
\eqno(7.10)
$$
leads to:
$$
\eqalign{
&\Sigma_i(\lambda,x,y)=a_i+b_iS\left[(\lambda
xy)^2,(x^2y)^{-1}\right],\quad i=1,2,\cr
&a_1=-(\lambda xy)^2{x^2-1\over x^4y^2-1},\cr
&b_1=1+(\lambda y)^{-2}{x^2y^2-1\over x^4y^2-1}+(\lambda
xy)^2{x^2-1\over x^4y^2-1},\cr
&a_2={(\lambda xy)^2\over x^4y^2-1},\cr
&b_2=1+{\lambda^{-2}x^2-(\lambda xy)^2\over x^4y^2-1}.\cr} 
\eqno(7.11)
$$
These relations, with $x=r/s$, $y=s$, and equation~(6.13), with $f_1=1$,
finally give 
$$
\eqalign{
&\Sigma(\lambda,r,s,u)=a+bS\left[\lambda^2r^2,{s\over 
r^2}\right],\cr
&a=\lambda^2r^2{s^2-r^2u^2\over r^4-s^2},\cr
&b=1+{\lambda^{-2}(r^2-u^2)+\lambda^2r^2(r^2u^2-s^2)\over r^4-s^2},\cr}
\eqno(7.12)
$$
which, together with (7.6), solves the site problem. Some
examples of the temperature variation of  the surface magnetization are
shown in figures~1 and~2. Similar
curves, for the bond problem, can be found in Turban \etal (Turban
\etal 1994). 

{\par\begingroup\medskip
\epsfxsize=9truecm
\topinsert
\centerline{\epsfbox{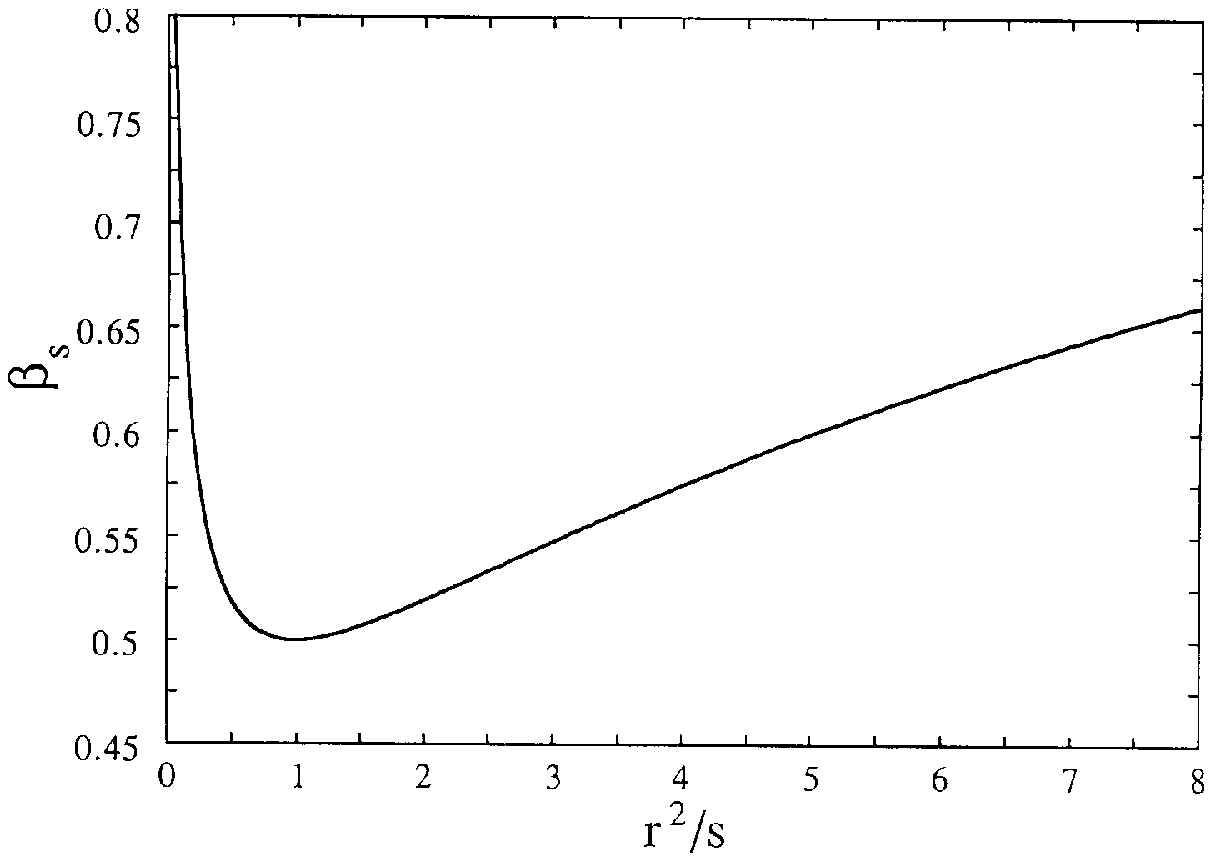}}
\smallskip
\figure{Variation of the surface magnetization  exponent with $r^2/s$ in the
site problem for the period--doubling sequence.}
\endinsert
\endgroup
\par}

The critical behaviour of the surface magnetization for
the site problem is governed by the singularity of 
$S(\lambda',r')$ in~(7.12), at $\lambda_c'$ given
by~(7.2), with $\lambda'\!=\!\lambda^2r^2$, $r'\!=\! sr^{-2}$. It
follows that the critical coupling satisfies $\lambda_c^2r^2\!=\!
(sr^{-2})^{-2/3}$, in agreement with~(7.8). Changing $\lambda_c$
into $\lambda_c'\!=\!(sr^{-2})^{-2/3}$ in equation~(7.7), one obtains
the surface magnetization exponent for the site problem
$$
\beta_s={1\over
2}{\ln\left[\left(r^2s^{-1}\right)^{1/3}+\left(r^2s^{-1}\right)^{-1/3}
\right]\over\ln 2} .  
\eqno(7.13)
$$
When $r\!=\!s$, the critical exponent is the same as for the bond problem
in~(7.7), as shown in section~6. The variation with
$r^2/s$ is shown in figure~3. One may notice that the value
$\beta_s\!=\! 1/2$, correspondind to the surface exponent of the
homogeneous two--dimensional Ising model at the ordinary surface transition,
is recovered when $s\!=\! uv\!=\!r^2$ and $\lambda_c\!=\! r^{-1}$. This
result is linked to the absence of $AA$ pairs in the sequence. This
value is also the minimum value of the exponent $\beta_s$, i.e. the
aperiodicity generally weakens the singularity.
\section{Thue--Morse sequence} 
As a second example, we consider the binary Thue--Morse
sequence (Dekking \etal 1983a), which, as mentioned in table~2,
leads to an irrelevant perturbation for the bond problem, treated in
Turban \etal (Turban \etal 1994), and to a marginal one for the site
problem. Starting with the letter $A$, the Thue--Morse subsitution
in~table~2 give, successively, 
$$
\eqalign{
&A\cr
&A\  B\cr 
&A\  B\  B\  A\cr
&A\  B\  B\  A\  B\  A\  A\  B\cr
&\underline{A}\  B\ \underline{B}\  A\ \underline{B}\  A\ 
 \underline{A}\  B\ \underline{B}\  A\ \underline{A}\  B\ 
 \underline{A}\  B\ \underline{B}\  A\cr
&\dots\cr}
\eqno(8.1) 
$$

In the bond problem, the critical coupling
$$
\lambda_c=r^{-1/2}\qquad {\rm (bond\  problem})
\eqno(8.2)
$$
follows from the asymptotic density
$\rho_\infty\!=\!1/2$ given by equation~(3.6).   
The form of the substitutions immediately leads to
the recursion relations 
$$
f_{2k}=1-f_k,\qquad f_{2k+1}=f_{k+1},
\eqno(8.3)
$$
which can be used in~(6.2) to give
$$
n_{2j}=j,\qquad n_{2j+1}=j+f_{j+1} .
\eqno(8.4)
$$
A chain with length $L\!=\!2j$ has a density equal to the asymptotic
one,  which explains the vanishing second
eigenvalue $\Lambda_2$ in table~2.

Since the calculation of $m_s$ has
already been described elsewhere (Turban \etal 1994), we
just mention the results. The surface magnetization follows from 
$$
\eqalign{
&S(\lambda,r)=\!{1\!+\! r\left(\lambda_c/\!\lambda\right)^2\over 
1\!-\!\left(\lambda_c/\!\lambda\right)^4}\!+\!(r^{-1}\!-\! r)\!
\left({\lambda\over\lambda_c}\right)^{\! 2}\!\! S_{TM}\!\left[
\!\left(\!{\lambda_c\over\lambda}\!\right)^{\! 4}\right],\cr
&S_{TM}(x)=x\sum_{k=0}^\infty{x^{2^k}\prod_{p=0}^k\left(1-x^{2^p}\right)
\over\left(1-x^{2^k}\right)\left(1-x^{2^{k+1}}\right)},\cr}
\eqno(8.5)
$$
where $S_{TM}(x)\!=\!\sum_{k=1}^\infty f_kx^k$ is the Thue--Morse
series (Dekking \etal 1983a). Near the critical point, one obtains
$$
m_s={2t^{1/2}\over\lambda_c+\lambda_c^{-1}}\left[1+{1\over 4}\left(
{1-\lambda_c^2\over 1+\lambda_c^2}\right)^2t+O(t^2)\right],
\eqno(8.6)
$$
where $t\!=\! 1-(\lambda_c/\lambda)^2$ is the deviation from the critical
point. The surface magnetization exponent takes its unperturbed value,
$\beta_s=1/2$, as expected for an irrelevant perturbation. 

In the site problem, with $\rho_\infty^{AA}\!=\!\rho_\infty^{BB}\!=\! 1/6$ 
and $\rho_\infty^{AB}\!=\!\rho_\infty^{BA}\!=\! 1/3$, the critical coupling
takes the form 
$$ 
\lambda_c=r^{-1/6}s^{-1/3}\qquad ({\rm site\  problem}).
\eqno(8.7)
$$
Making use of~(8.3), equation~(6.7) leads to 
$$
m_{2j}=m_{2j+1}=n_{j+1}-m_j.
\eqno(8.8)
$$
The same result is obtained for even and odd terms because the two letters
occur within successive pairs along the sequence, so that
$f_{2k+1}f_{2k+2}\!=\! 0$.

With a sequence starting on $A$, the front factor disappears in the sum
$\Sigma(\lambda,r,s,u)$ of equation~(6.13).
The $\Sigma_i$s, defined in~(6.14), satisfy the functional
equations 
$$
\eqalign{
&\Sigma_i(\lambda,x,y)=\sum_{j=1,2}a_{ij}\Sigma_j(\lambda^2y,x^{-1},x),
\quad i=1,2,\cr
&a_{11}=1+\lambda^{-2},\cr
&a_{12}=x^{-2}-\lambda^{-2}+(\lambda xy)^{-2}-1,\hfil\cr
&a_{21}=\lambda^{-2},\cr
&a_{22}=x^{-2}-\lambda^{-2},\hfil\cr}
\eqno(8.9)
$$
which, as usual, are obtained by splitting the sums into even and odd
parts and using~(8.4), (8.8), as well as the
identity~(6.12).

{\par\begingroup\medskip
\epsfxsize=9truecm
\topinsert
\centerline{\epsfbox{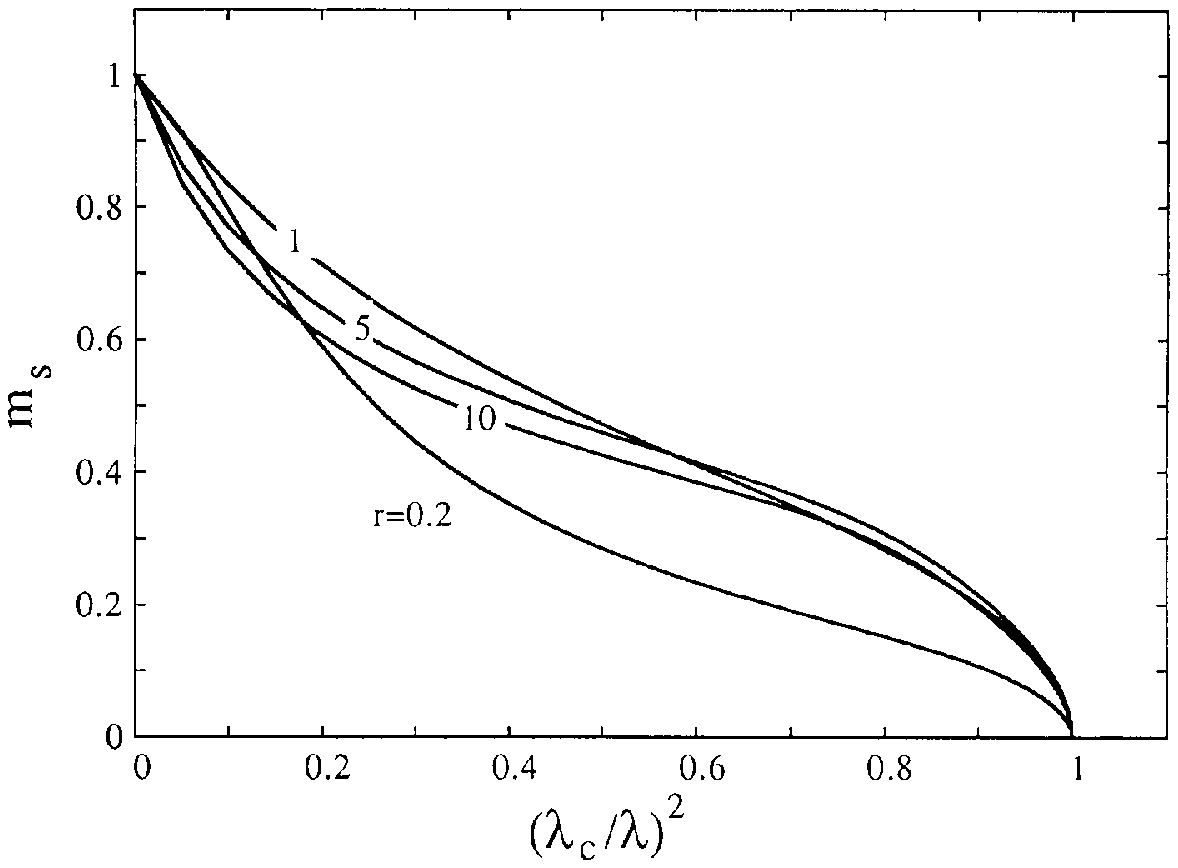}}
\smallskip
\figure{Temperature dependence of the surface
magnetization (Thue--Morse, site problem) for different values of
$r$, $u\!=\!.5$ and $s\!=\!1$.}
\endinsert
\endgroup
\par}

{\par\begingroup\medskip
\epsfxsize=9truecm
\topinsert
\centerline{\epsfbox{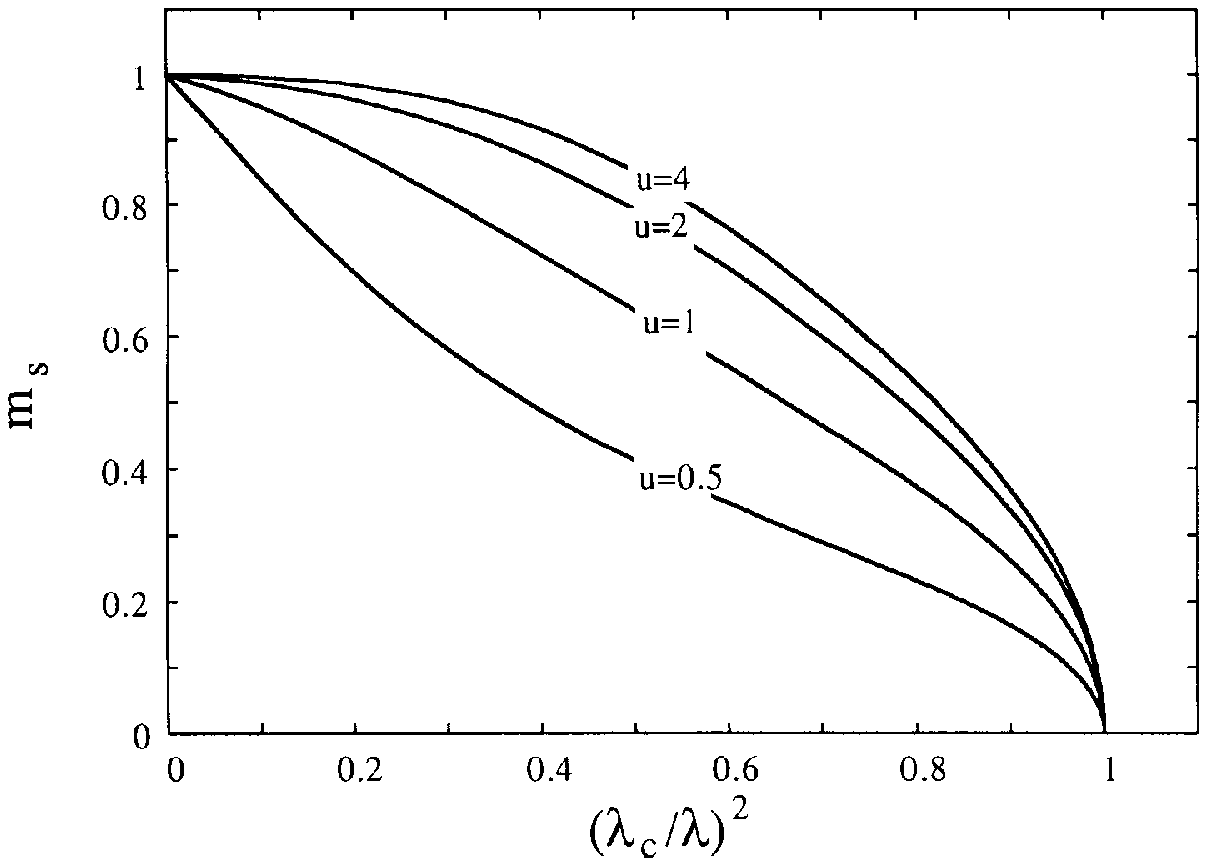}}
\smallskip
\figure{Temperature dependence of the surface
magnetization (Thue--Morse, site problem) for different values of
$u$, $r\!=\!.5$ and $s\!=\!1$.}
\endinsert
\endgroup
\par}

Through iteration, at step $k\geq 1$, the arguments of the $\Sigma_i$s
become 
$$
\lambda_k=\lambda^{2^k}x^{[2^{k-1}+(-1)^k]/3}y^{2^{k-1}},\quad
x_k=y_k^{-1}=x^{(-1)^k}
\eqno(8.10)
$$
and the functional equations~(8.9) can be rewritten in matrix
form as 
$$
\left(\matrix{\Sigma_1(\lambda_k,x_k,y_k)\cr
              \Sigma_2(\lambda_k,x_k,y_k)\cr}\right)=
{\bss T_k}
\left(\matrix{\Sigma_1(\lambda_{k+1},x_{k+1},y_{k+1})\cr
              \Sigma_2(\lambda_{k+1},x_{k+1},y_{k+1})\cr}\right),  
\eqno(8.11)
$$
with
$$
{\bss T_k}=\left(\matrix{
               1+\lambda_k^{-2}&x_k^{-2}-1\cr
               \lambda_k^{-2}&x_k^{-2}-\lambda_k^{-2}\cr}\right),
\qquad k\geq 1.
\eqno(8.12)
$$
Then,
$$
\left(\matrix{\Sigma_1(\lambda_1,x_1,y_1)\cr
              \Sigma_2(\lambda_1,x_1,y_1)\cr}\right)=\prod_{k=1}^\infty
{\bss T_k}\left(\matrix{1\cr 0\cr}\right),  
\eqno(8.13)
$$
where the components of the vector on the right follows from the form of
the $\Sigma_i$s when $k\!\to\!\infty$. 

Equations~(6.11), (6.13), and (8.11--13),
formally solve the problem. The \discretionary{tempe-}{rature}{temperature}
variation of the surface magnetization is shown in figures~4 and~5.

It seems difficult to obtain an explicit expression for the surface
magnetization since the form of the matrix ${\bss T_k}$ depends on the index
$k$. But some progress can be made concerning the critical behaviour, using
the same scaling method as for the period doubling sequence, i.e., looking at
the $L$-dependence of the truncated series $\Sigma_L(z)$ at the critical
point $z\!=\!(\lambda_c/\lambda)^2\!=\! 1$.  

{\par\begingroup\medskip
\epsfxsize=9truecm
\topinsert
\centerline{\epsfbox{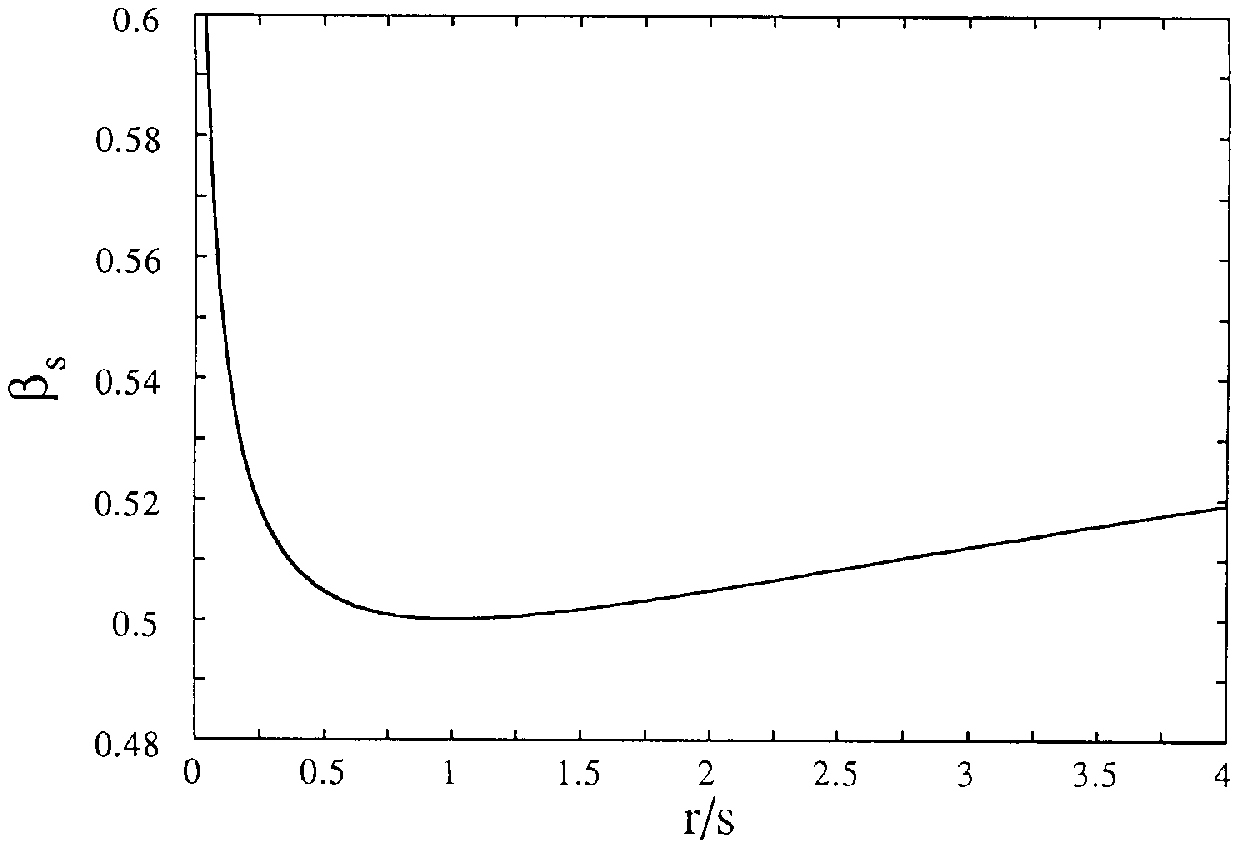}}
\smallskip
\figure{Variation of the surface magnetization
exponent with $r/s$ in the site problem for the Thue--Morse
sequence.}
\endinsert
\endgroup
\par}

With the actual values of the arguments
in~(6.13),  $x=r/s$ and $y=s$, equation~(8.10) gives
$$
\lambda_k=\left({r\over s}\right)^{(-1)^k/3}
\left({\lambda\over\lambda_c}\right)^{2^k},\quad
x_k=y_k^{-1}=\left({r\over s}\right)^{(-1)^k},
\eqno(8.14)
$$
so that, at the critical point, the matrix elements in ${\bss T_k}$
depend on $k$ only through $(-1)^k$. Let us introduce the product 
${\bss U}\!=\!{\bss T_{2p-1}}\ {\bss T_{2p}}$ with
$$
{\bss U}=\left(\matrix{
    2+w+w^2&(w^{-1}-w)(1+w+w^{-1}\cr
    w+w^2&(1-w)(2+w)\cr}\right),\quad
w=\left({r\over s}\right)^{2/3} .
\eqno(8.15)
$$
Taking $L\!=\!2^{2l}$, the first $L$ terms in $\Sigma_L(z)$ are obtained,
up to  $L$--independent factors, keeping the first $2l$ terms in the
infinite product of equation~(8.13). Due to the form of the matrix elements
in~(8.12), they belong to the first column of the matrix resulting
from the finite product. Thus, using~(8.15), they are also given
by ${\bss U}^l\left(\matrix{1\cr 0\cr}\right)$ at the critical point.
After diagonalization, one obtains
$$
\Sigma_L(1)\sim (2^{2l})^{2\beta_s}\sim\left(w^{1/4}+w^{-1/4}\right)^{2l}
\eqno(8.16)
$$
where the last term is the $l$th power of the largest eigenvalue of
${\bss U}$. Finally, the surface magnetization exponent reads
$$
\beta_s={1\over
2}{\ln\left[(rs^{-1})^{1/6}+(rs^{-1})^{-1/6}\right]\over\ln2} .
\eqno(8.17) 
$$ 
The surface magnetization exponent is shown as a function of $r/s$ in
figure~6. The minimum value $\beta_s\!=\! 1/2$,
corresponding to the homogeneous system and to the bond problem behaviour
as well, is reached when $r\!=\! s$. 

\section{Conclusion} 
We have presented a comparative study of the influence of bond and site
aperiodicities on the critical behaviour at a second order phase
transition. One dimensional sequences generated
through substitution, corresponding to an uniaxial aperiodic modulation,
have been discussed. The extension of the Harris criterion to the 
site problem requires the knowledge of the eigenvalues of a substitution
matrix which is linked to the distribution of pairs of successive letters
along the sequence. For sequences with $\vert\Lambda_2\vert\!\geq\!1$, the
relevance--irrelevance criterion is the same in the bond and site problems.
When $\vert\Lambda_2\vert\!<\!1$, the aperiodic perturbation may become
more dangerous in the site problem, depending on the form of the
substitution. 

Exact results have been obtained for the surface magnetization of layered
Ising aperiodic systems. The period--doubling sequence leads to a marginal
perturbation for both problems, but with different nonuniversal exponents.
The Thue--Morse aperiodic modulation, which is irrelevant for the bond
problem, becomes marginal for the site problem, where the surface magnetic
exponent is nonuniversal. 

Among the possible extensions of this work, one may mention the 
treatment of substitutions with more than two letters and,
more interesting, the study of higher--dimensional aperiodic
perturbations in isotropic or anisotropic critical systems.

\ack
LT and BB are indebted to Ferenc Igl\'oi for
stimulating discussions and collaboration in this, and related fields.
This work was supported by CNIMAT under project No 155C93b.

\references

\refjl{Benza G V 1989}{Europhys. Lett.}{8}{321}

\refbk{Collet P and Eckmann J P 1980}{Iterated Maps on
the Interval as Dynamical Systems}{(Boston: Birkh\" auser)}

\refjl{Dekking M, Mend\`es--France M and van der
Poorten A 1983}{Math. Intelligencer}{4}{130}

\refjl{\dash 1983}{Math. Intelligencer}{4}{190}

\refjl{Doria M M and Satija I I 1988}{\PRL}{60}{444}

\refbk{Dumont J M 1990}{Number Theory and Physics, Springer Proc. Phys.}
{vol 47 ed. J M Luck, P Moussa and  M Waldschmidt (Berlin: Springer) p 185}

\refjl{Garg A and Levine D 1987}{\PRL}{59}{1683} 

\refjl{Godr\`eche C, Luck J M and Orland H J 1986}{J. Stat. Phys.}{45}{777} 

\refjl{Godr\`eche C and Lan\c con F 1992}{\JP\ {\rm I}}{2}{207}

\refjl{Grimm U and Baake M 1994}{J. Stat. Phys.}{74}{1233}

\refjl{Guyot P, Kramer P, and de Boissieu M 1991}{\RPP}{54}{1373} 

\refjl{Harris AB 1974}{\JPC}{7}{1671}

\refjl{Henkel M and Patk\'os A 1992}{\JPA}{25}{5223}

\refjl{Henley C L 1987}{Comments Cond. Matter Phys.}{13}{59}

\refjl{\dash and Lipowsky R 1987}{\PRL}{59}{1679} 

\refjl{Igl\'oi F 1986}{\JPA}{19}{3077}

\refjl{\dash 1988}{\JPA}{21}{L911}

\refjl{\dash 1993}{\JPA}{26}{L703}

\refjl{\dash and Turban L 1994}{Europhys. Lett.}{27}{91}

\refjl{Janot C, Dubois J M and de Boissieu M 1989}{Am. J. Phys.}{57}{972} 

\refjl{Janssen T 1988}{Phys. Rep.}{168}{55}

\refjl{Jordan P and Wigner E 1928}{\ZP}{47}{631}

\refjl{Kogut J 1979}{\RMP}{51}{659} 

\refjl{Langie G and Igl\'oi F 1992}{\JPA}{25}{L487} 

\refjl{Lieb E H, Schultz T D and Mattis D C 1961}{\APNY}{16}{406}

\refjl{Lin Z and Tao R 1992}{\JPA}{25}{2483}

\refjl{Luck J M 1993a}{J. Stat. Phys.}{72}{417}

\refjl{\dash 1993b}{Europhys. Lett.}{24}{359}

\refjl{McCoy  B M and Wu T T 1968a}{\PRL}{21}{549} 

\refjl{\dash 1968b}{\PR}{176}{631}

\refjl{McCoy B M 1970}{\PR\ {\rm B}}{2}{2795}

\refjl{Okabe Y and Niizeki K 1988}{\JPSJ}{57}{1536}

\refjl{\dash 1990}{\JPA}{23}{L733}

\refjl{Peschel I 1984}{\PR\ {\rm B}}{30}{6783} 

\refjl{Pfeuty P 1979}{\PL}{72A}{245}

\refbk{Queff\'elec M 1987}{Substitution Dynamical
Systems--Spectral Analysis, Lecture Notes in Mathematics}{vol 1294
ed. A Dold and B Eckmann (Berlin: Springer) p 97}

\refjl{Sakamoto S, Yonezawa F, Aoki K, Nos\'e S and 
Hori M  1989}{\JPA}{22}{L705}

\refjl{Shechtman D, Blech I, Gratias D and Cahn J W 1984}{\PRL}{53}{1951}

\refjl{Schultz T D, Mattis D C and Lieb E H 1964}{\RMP}{36}{856}

\refjl{S\o rensen E S, Jari\'c M V and Ronchetti M 1991}{\PR\ 
{\rm B}}{44}{9271}

\refbk{Steinhardt P and DiVicenzo D 1991}{Quasicrystals: 
The State of the Art}{ed.~P.~Steinhardt and 
D. DiVicenzo (Singapore: World Scientific)}

\refjl{Tracy C A 1988}{\JPA}{21}{L603}

\refjl{Turban L and Berche B 1993}{\ZP\ {\rm B}}{92}{307}

\refjl{Turban L, Igl\'oi F and Berche B 1994}{\PR\ {\rm B}}{49}{12695}

\refjl{Zhang C and De'Bell K 1993}{\PR\ {\rm B}}{47}{8558} 

\vfill\eject\bye